\documentclass[twocolumn]{autart}    % Enable this line and disable the
\usepackage{latexsym,amssymb}
\usepackage{mathrsfs}
\usepackage{epsfig}
\usepackage{graphicx}
\usepackage{color}
\usepackage{enumerate}
\usepackage{amsmath,amssymb,amsfonts}
\usepackage{breqn}
\usepackage{bm}
\usepackage{subcaption}
\usepackage{algorithm} %format of the algorithm
\usepackage{algorithmic} %format of the algorithm

\newtheorem{theorem}{Theorem}

\newtheorem{proposition}{Proposition}
\newtheorem{assumption}{Assumption}

\newtheorem{remark}{Remark}
\newenvironment{definition}{\medskip\noindent{\it Definition. }}{ \medskip}
\newcommand{\R}{\mathbb{R}}

\DeclareFontFamily{OT1}{pzc}{}
\DeclareFontShape{OT1}{pzc}{m}{it}{<-> s * [1.200] pzcmi7t}{}
\DeclareMathAlphabet{\mathpzc}{OT1}{pzc}{m}{it}

\newcommand*\mcapinn[2]{\vcenter{\hbox{$\mathsurround=0pt
  \ifx\displaystyle#1\textstyle\else#1\fi\bigcap$}}}

\newcommand*\mcupinn[2]{\vcenter{\hbox{$\mathsurround=0pt
  \ifx\displaystyle#1\textstyle\else#1\fi\bigcup$}}}

\def\begequarr{\begin{eqnarray}}
\def\endequarr{\end{eqnarray}}
\def\begequarrs{\begin{eqnarray*}}
\def\endequarrs{\end{eqnarray*}}
\def\begequ{\begin{equation}}
\def\endequ{\end{equation}}
\def\begequs{\begin{equation*}}
\def\endequs{\end{equation*}}
\def\begite{\begin{itemize}}
\def\endite{\end{itemize}}

\def\begcen{\begin{center}}
\def\endcen{\end{center}}
\def\begrem{\begin{remark}\rm}
\def\endrem{\end{remark}}
\def\ba{\begin{array}}
\def\ea{\end{array}}

\def\diag{\textnormal{diag}}

\def\col{\textnormal{col}\; }

\def\calL{\mathcal{L}}
\def\calN{\mathcal{N}}

\def\col{\textnormal{col}}

\newcommand{\db}{\mathbf{d}}

\newcommand{\mV}{\mathrm{V}}
\newcommand{\mG}{\mathrm{G}}
\newcommand{\mE}{\mathrm{E}}
\newcommand{\mN}{\mathrm{N}}

\newcommand{\ub}{\mathbf{u}}

\newcommand{\xb}{\mathbf{x}}

\newcommand{\etab}{\bm{\eta}}

\newcommand{\Lb}{\mathbf{L}}
\newcommand{\Ib}{\mathbf{I}}

\newcommand{\Qb}{\mathbf{Q}}

\newcommand{\Pb}{\mathbf{P}}

\newcommand{\Ab}{\mathbf{A}}

\newcommand{\eb}{\mathbf{e}}

\newcommand{\Ub}{\mathbf{U}}

\newcommand{\omegab}{\bm{\omega}}
\newcommand{\xib}{\bm{\xi}}

\newcommand{\gammab}{\bm{\gamma}}

\begin{document}

\begin{frontmatter}

\title{Differentially Private Average Consensus with  Improved Accuracy-Privacy Trade-off} % Title, preferably not more
                                                % than 10 words.
\vspace{-2.5em}

\author[ZJU]{Lei Wang}\ead{lei.wangzju@zju.edu.cn},
\author[ZJU]{Weijia Liu}\ead{liuweijia0430@163.com},
\author[ZJUT]{Fanghong Guo}\ead{fhguo@zjut.edu.cn},
\author[SHICT]{Zixin Qiao}\ead{qiaozixin2021@outlook.com},
\author[ZJU]{Zhengguang Wu}\ead{nashwzhg@zju.edu.cn}

\address[ZJU]{College of Control Science and Engineering, Zhejiang University, Hangzhou, China.}
\address[ZJUT]{Department of Automation, Zhejiang University of Technology, Hangzhou, China.}
\address[SHICT]{Shanghai Institute of Computing Technology, Shanghai, China.}
\thanks{This paper was partially supported by  the National Key R\&D Program of China under Grant 2018YFA0703800, the National Natural Science Foundation of China under Grant No. 62203386, 62373328,  and Zhejiang Provincial Natural Science Foundation of China under Grant No. LZ23F030008. (Corresponding author: Fanghong Guo)}
\thanks{A preliminary work has been presented at 2022 34th Chinese Control and Decision Conference (CCDC) \cite{Qiao2022CCDC}.}

\vspace{-1.5em}

\vspace{-0.5em}
\begin{abstract}                          % Abstract of not more 
This paper studies the average consensus problem with differential privacy of initial states, for which it is widely recognized that there is a trade-off between the mean-square computation accuracy and privacy level.  
Considering the trade-off gap between the  average consensus algorithm and the centralized averaging approach with differential privacy, we propose a distributed shuffling mechanism based on the Paillier cryptosystem to generate correlated zero-sum randomness. 
By randomizing each local privacy-sensitive initial state with an i.i.d. Gaussian noise and the output of the mechanism using Gaussian noises, it is shown that the resulting average consensus algorithm can eliminate the gap in the sense that the accuracy-privacy trade-off  of the centralized averaging approach with differential privacy can be almost recovered by appropriately designing the variances of the added noises. We also extend such a design framework with Gaussian noises to the one using Laplace noises, and show that the improved privacy-accuracy trade-off is preserved. 
\end{abstract}

\end{frontmatter}

%%%%%%%%%%%%%%%%%%%%%%%%%%%%%%%%%%%%%%
%%%%%%%%%%%%%%%%%%%%%%%%%%%%%%%%%%%%%
\section{Introduction}
\vspace{-1em}
With the  development of cyber-physical systems in such as smart grids and intelligent transportation, it becomes an emerging problem to complete a computation task defined by a network of physically distributed agents, i.e., each holds a part of the task and expects to solve it cooperatively by communicating with neighboring agents over the network. 
In distributed computations,  the consensus algorithm plays a fundamental role of 
acting as a standard information aggregation routine, e.g., in solving distributed optimization \cite{nedic2018distributed,yang2019survey} and  network linear equations \cite{CPSLinearEqu1,CPSLinearEqu2}.
Among consensus algorithms, the average consensus is the most basic one with wide applications in practical fields, such as modeling ocean temperature with mobile sensors \cite{0201EnvironmentalModeling} and distributed load shedding in smart grids \cite{mortaji2017load,Qiao2022CCDC}. 

\vspace{-1em}
In the standard average consensus algorithm, each local data for average computation is assigned as the initial state of the local agent, which  communicates with neighboring agents the agent state for state iterations until reaching convergence (to the average).
Note that the network data for average computation may be privacy-sensitive. For example, when applying the average consensus to solve the load shedding problem, the network data contain the local bus power demands, which may represent activities of  the corresponding family and thus may be sensitive and should be protected from adversaries. 
However, when computing the average, the adversaries may have access to the communication messages and  infer the network data under some observability condition, leading to privacy leakage risks.

\vspace{-1em}
To achieve the privacy-preserving average consensus, a natural idea is to apply the encryption-based techniques to protect the sensitive data or communication messages such that the eavesdropped information is hardly used to infer the sensitive data \cite{3102PPOverview,ruan2019secure,hung2023}. Particularly, in \cite{ruan2019secure} the Paillier cryptosystem is employed to develop secure average consensus algorithms such that the average computation is completed in ciphertexts, i.e., there is no need to use the private key for decryption during computation. 
% {\color{blue} Along using Paillier cryptosystem, \cite{hung2023} proposed a distributed manner to achieve exact average consensus while preserving agent's privacy even when all of its neighbors are untrusted.}
Though providing privacy and accuracy guarantees simultaneously, the computational and communication costs for encryption-based algorithms may be too heavy in practical applications \cite{3101RealTimeMulAgentsChallenge}.

\vspace{-1em}
Another common approach for privacy protection is to add offsets or masks to node states or their iteration processes \cite{IniValObser1,IniValObser2,3202PPAC,farokhi2019ensuring}. Along this line,  \cite{IniValObser1} proposed to add  offsets in such a way that for each node locally added offsets are zero in total, which ensures the exact average computation while achieving the privacy in the unobservability sense. Similar results have also been achieved in \cite{IniValObser2} by introducing time-varying output masks such that the masked time-varying system has the original system as its limit system.  To quantify the achievable privacy, the variance matrix of {the maximum likelihood estimation} was employed in  \cite{3202PPAC}, where a privacy-preserving average consensus algorithm was established by adding and subtracting vanishing random noises. From a different perspective, \cite{farokhi2019ensuring} proposed to add constrained noises in  consensus processes, where {the inverse of the trace of the Fisher information matrix was used to measure the privacy guarantee}.
{Note that in these efforts the privacy guarantees are established on the eavesdropped/accessed information, with the effect of side information not considered.} 
%{\color{blue} Compared with differential privacy based methods,} these noise adding methods hold uncertain robustness to the side information. 

\vspace{-1em}
Differential privacy,  a rigorous notion for defining and preserving data privacy, has been shown to be resilient to the side information and post-processing \cite{dwork2006calibrating,dwork2014algorithmic}. In last decades, extensive developments have been emerged in such as signal processing \cite{DPScenarioMSS,le2013differentially}, control \cite{kawano2020design,kawano2021modular,yazdani2022differentially}, and distributed computation \cite{han2016differentially,nozari2016differentially}, etc, advancing the differential privacy as a gold standard in data privacy. 
Taking into account the average consensus with differential privacy of initial states, there are also many efforts in the literature \cite{3201DPIterative,nozari2017differentially,liu2020differentially,wang2023differential,he2020differential,yqwang2023robustDAC}.  Particularly, \cite{3201DPIterative} developed an iterative consensus framework by adding a stream of noise drawn from a time-varying Laplace distribution. In \cite{nozari2017differentially}, a differentially private  consensus algorithm was proposed by linearly perturbing the state-iterating processes and communicated messages with exponentially decaying  noise. It is also shown in \cite{nozari2017differentially} that given adversaries having access to all the messages, achieving exact average in the mean-square sense is \emph{impossible} for average consensus algorithms under the requirement of differential privacy of the agents' initial states, and the corresponding  optimal trade-off between the computation accuracy and the differential privacy can be achieved by the mechanism corresponding to the  one-shot perturbation of initial states. 
With such an optimal trade-off,  it is worth noting that  for differential privacy, the centralized average mechanism (i.e., publishing the perturbed average) shows a better accuracy-privacy trade-off, as shown in Section \ref{sec-2-B}. More explicitly, given the same differential privacy requirement, the mean-square computation accuracy that can be achieved by the centralized approach is $n$ times smaller than  that of the  average consensus algorithm with the  one-shot perturbation of initial states for differential privacy \cite{nozari2017differentially}, where $n$ denotes the total agent number.

\vspace{-1em}
Motivated by the previously mentioned gap between the centralized approach and the average consensus algorithms in the literature for differential privacy, in this paper we revisit the average consensus problem with the requirement of differential privacy of agents' initial states against adversaries having access to all the messages, and aim to propose new differentially private average consensus algorithms with improved accuracy-privacy trade-offs. {On the other hand, for the problem of computing the average of \emph{dynamic} network data, \cite{yqwang2023robustDAC}  proposed to modify the adjacency  when characterizing the differential privacy of the dynamic data, and then introduced a decaying factor to attenuate the influence of noise. As a result, an exact average consensus can be computed while achieving the desired privacy requirement. However, due to the specifically modified adjacency notion, the resulting algorithm in \cite{yqwang2023robustDAC} cannot be applied to handle the  problem considered in the paper where a network of static data is concerned and the differential privacy notion follows the conventional one \cite{le2013differentially,han2016differentially,nozari2016differentially}.}

\vspace{-1em}
In this paper, inspired by \cite{ruan2019secure}, we propose a distributed  shuffling mechanism based on the Paillier cryptosystem to generate correlated zero-sum randomness. With such a mechanism using Gaussian noises, we then inject the resulting correlated randomness and an \emph{extra} i.i.d. Gaussian noise to  the local data for average computation as the initialization step of the average consensus algorithm. 
It is shown that the resulting average consensus algorithm can preserve the desired differential privacy, while achieving exponential convergence to the average subject to an error relying on the added noises. We also extend such a design framework with Gaussian noises to the one using Laplace noises. 
Our contribution mainly lies in proposing two new design frameworks of differentially private average consensus algorithms, respectively, using Gaussian and Laplace noises, both of which can almost recover the accuracy-privacy trade-off of the corresponding centralized averaging approach. 
More explicitly, we show that, with the introduction of the proposed distributed shuffling mechanism and an extra i.i.d. Gaussian/Laplace noise in the initialization step, the resulting average consensus algorithms can eliminate the gap in the sense that the achieved trade-off can be adjusted arbitrarily close to that of the centralized averaging approach  by appropriately designing the variance of the added noises. 
%{\color{blue} Furthermore, our algorithms can thus be applied to solve other distributed computation problems such as the distributed least-squares optimization problem.}

\vspace{-1em}
The remainder of the paper is organized as follows. Section 2 presents the gap of the accuracy-privacy trade-off between the existing average consensus algorithm and the centralized algorithm for differential privacy, and formulates the problem of interest. In Section 3, the Paillier cryptosystem is employed to develop a distributed shuffling mechanism, which is then used in the initialization step of the average consensus algorithm in Sections 4 and 5 with Gaussian and Laplace noises, respectively for differentially private average consensus algorithms. Case studies are given in Section 6 to validate the effectiveness of the proposed algorithms. The conclusion is drawn in Section 7. This paper is a
significant extension over the preliminary version \cite{Qiao2022CCDC} by reformulating the problem in Section 2, developing new technical results in Section 5 and simulations in Section 6.

\noindent{\bf Notation}. Denote by $\R$ the real numbers, $\R^n$  the real space of $n$ dimension for any positive integer $n$ and $\mathbb{N}$ the set of natural numbers. For a vector $\xb\in\R^n$,  denote $x_i$ as the $i$-th entry of $\xb$, and $\|\xb\|_0$, $\|\xb\|_1$ and $\|\xb\|$ as the $0$, $1$, and $2$-norm of vector $\xb$, respectively.
Denote $\eb_i$ a basis vector whose entries are all zero except the $i$-th being one. 
Denote by $\etab\sim\mathcal{N}({\upsilon},\sigma^2)^r$ if each entry in $\etab\in\mathbb{R}^r$ is i.i.d. drawn from a Gaussian distribution with mean {$\upsilon$} and variance $\sigma^2$, and $\etab\sim\mathcal{L}({\upsilon},b)^r$, if each entry in $\etab\in\mathbb{R}^r$ is i.i.d. drawn from a Laplace distribution with mean ${\upsilon}$ and variance $2b^2$. 
{
For a matrix $\Lb$, denote $\|\Lb\|_1$ and $\|\Lb\|$ as its $1$- and $2$-norm, respectively, and $[\Lb]_{ij}$ as the element at the $i$-th row and the $j$-th column.
}
{For two functions $f,g$, we say $f(x)=\mathcal{O}(g(x))$ if there exists real numbers $M_0>0$ and $x_0$ such that $|f(x)|\leq M_0 g(x)$ for all $x\geq x_0$.
For a matrix $U\in\mathbb{R}^{m\times n}$ and a set $\mathcal{M}\subset\mathbb{R}^n$, we define $U\mathcal{M} :=\{\textbf{x}|\textbf{x}=U\textbf{m}, \textbf{m}\in\mathcal{M}\}$.
}
Define $\Phi(s):=\frac{1}{\sqrt{2\pi}}\int_{-\infty}^s e^{-\tau^2/2}d\tau$ and 
$
\kappa_{\epsilon}(s):= \Phi(\frac{s}{2} - \frac{\epsilon}{s}) - e^{\epsilon}\Phi(-\frac{s}{2} - \frac{\epsilon}{s})
$.
Denote ${\kappa^{-1}_\epsilon(\delta)}$ as the inverse function\footnote{It can be verified that  $\frac{\partial \kappa_{\epsilon}(s)}{\partial s}=  e^{-\frac{1}{2}(\frac{s}{2} - \frac{\epsilon}{s})^2} /{\sqrt{2\pi}} >0$. This indicates that the function $\kappa_{\epsilon}(\cdot)$ and thus its inverse ${\kappa^{-1}_\epsilon(\cdot)}$ are strictly increasing functions.} of $\kappa_{\epsilon}(s)$ for any $\epsilon\geq 0$, i.e., ${\kappa^{-1}_\epsilon}(\kappa_{\epsilon}(s))=\kappa_{\epsilon}({\kappa^{-1}_\epsilon}(s))=s$ for $s>0$.

\vspace{-1em}
\section{Problem Statement}
\vspace{-1em}
\subsection{Preliminaries}
\vspace{-1em}

In this paper, we study the problem of average consensus over a communication network $\mG=(\mV,\mE)$ where $\mathrm{V}=\{1,\ldots,n\}$ is the agent set, $\mathrm{E}\subseteq\mathrm{V}\times\mathrm{V}$ is the edge set, and each agent $i\in\mV$ holds a \emph{privacy-sensitive} local {$d_i\in\mathbb{R}$}. {Note that the data $d_i$ is assumed to be scalar for simplicity, though the extension to a vector $\textbf{d}_i$ can be easily achieved by appropriately adapting the forthcoming node state and communication message to be vectors and the generated/added randomness to be vectors of i.i.d. noises.} Throughout the paper, the following assumption on the communication network $\mG$ is made.

\vspace{0.5em}
\begin{assumption}\label{ass-1}
The communication graph $\mG$ is undirected and connected. Moreover, denote by $w_{ij}$  the weight of the edge $(i,j)$, satisfying {$w_{ii}=0$ for all $i\in\mathrm{V}$}, $w_{ij}=w_{ji}>0$ if $(i,j)\in\mathrm{E}$, $w_{ji}=0$ if $(i,j)\notin\mathrm{E}$, and $\sum_{j\in\mathrm{V}}w_{ij} <1$ for all $i\in\mathrm{V}$.
\end{assumption}

Denote the Laplacian matrix of $\mG$ as $\Lb$, satisfying $[\Lb]_{ij}=-w_{ij}$, $j\neq i$ and $[\Lb]_{ii}=\sum_{k=1}^n w_{ik}$ for all $i\in\mV$.
Let us arrange the eigenvalues of $\Lb$ in the increasing order as $\lambda_1^{\Lb} \leq \lambda_2^{\Lb}\leq ...\leq \lambda_n^{\Lb}$. By \cite[Theorem 2.8]{mesbahi2010graph} {and Gershgorin Circle Theorem \cite{horn_johnson_1985}}, Assumption 1 implies $0=\lambda_1^{\Lb}<\lambda_2^{\Lb}\leq\lambda_n^{\Lb}<2$. Denote  $\beta=\max\{|1-\lambda_2^{\Lb}|,|1-\lambda_n^{\Lb}|\}$, satisfying $0\leq\beta <1$.

\vspace{-1em}
Denote the neighboring set of agent $i\in\mV$ as $\mN_i$. A standard algorithm to solve the average consensus problem {\cite{mesbahi2010graph}} follows
\begin{equation}\label{eq:AveCon}
  x_i(t+1) = x_i(t) + \sum_{j\in\mN_i}w_{ij}(x_j(t) - x_i(t))\,.
  \vspace{-1em}
\end{equation}
By initializing each agent state $x_i(0)=d_i$, it is well-known that under Assumption \ref{ass-1}, each agent state $x_i(t)$ converges to the average $d^\ast:=\sum_{i=1}^n d_i/n$ exponentially  \cite{mesbahi2010graph}. However, it is worth noting that during the average computation, there may be adversaries who have access to the communication messages over the communication graph $\rm G$ and may infer the privacy-sensitive local data of the network. This thus leads to the study of modifying the algorithm \eqref{eq:AveCon} for privacy-preserving purpose. 
In view of this, this paper takes into account the differential privacy of these local data $d_i$, and aim to  develop new distributed average consensus algorithms to compute the average with privacy guarantees. 

\vspace{-1em}
Denote $\mathscr{M}:\mathcal{D}\rightarrow\mathcal{M}$ as the mapping from  the local data to the eavesdropped messages. Let $\mathcal{D}\subseteq\mathbb{R}^n$ be the input space of the private local data, and any pair of data $\xb$ and $\xb'$ drawn from  $\mathcal{D}$ are said to be $\mu$-adjacent with $\mu>0$, denoted by $(\xb,\xb')\in\textnormal{Adj}(\mu)$, if $\|\xb-\xb'\|_0= 1$ and  $\|\xb-\xb'\|_1\leq \mu$.
We present the following definition  \cite{dwork2006calibrating}.

\vspace{-1em}
\begin{definition}\label{def:DP-1}
Denote the vector of the sensitive local data as $\db=[d_1;d_2;\ldots;d_n]$. A distributed average consensus algorithm over the communication graph $\mG$ preserves  $(\epsilon,\delta)$-differential privacy of $\db$ under $\mu$-adjacency for $\epsilon\geq0,\delta\in[0,1)$,  if  for all $\mathcal{M}\subseteq\mbox{range}(\mathscr{M})$, there holds
\begin{equation}\label{defeq:DP}
    \mathbb{P}\big(\mathscr{M}(\db) \in \mathcal{M} \big) \leq e^\epsilon \mathbb{P}\big(\mathscr{M}(\db^\prime) \in \mathcal{M} \big) +\delta\,
\end{equation}
for any $(\db,\db')\in\textnormal{Adj}(\mu)$. {If $\delta=0$, the resulting $(\epsilon,0)$-differential privacy  is usually called $\epsilon$-differential privacy.}
\end{definition}

% \vspace{1em}
{
\begin{remark}
In the literature both notions of $\epsilon$- and $(\epsilon,\delta)$-differential privacy are widely used, with the latter being a relaxed version of the former. To achieve $\epsilon$-differential privacy, a common idea is to inject Laplace noise with certain variance, yielding the Laplace mechanism and privacy budget $\epsilon>0$ \cite{le2013differentially}. For $(\epsilon,\delta)$-differential privacy, Gaussian noise is usually injected instead, yielding the Gaussian mechanism and privacy budget $\epsilon\geq0$ and $\delta\in(0,1)$ \cite{balle2018improving}. 
Thus, in this paper we focus on both  Gaussian  and Laplace mechanisms representative for $\epsilon$- and $(\epsilon,\delta)$-differential privacy, respectively. See Remark \ref{rem-ext} for possible extension to the use of other kinds of noises.
\end{remark}
}

\vspace{-1em}
\subsection{Problem Definition}
\label{sec-2-B}
\vspace{-1em}
Differentially private average consensus algorithms have been  investigated in  the literature (e.g. \cite{nozari2017differentially,liu2020differentially}).
Particularly,  it is shown in \cite{nozari2017differentially} that the optimal trade-off between the privacy and the accuracy is achieved by injecting an i.i.d. noise to each local data before assigning it as the initial state, i.e., the so-called one-shot perturbation following
\begin{equation}\label{eq:one-shot}
x_i(0)=d_i+\xi_i
\vspace{-1em}
\end{equation}
with the injected i.i.d. noise $\xi_i$,
and then running the standard average consensus algorithm \eqref{eq:AveCon}. For convenience, we call the above Differentially Private Average Consensus algorithm with the One-Shot Perturbation (\ref{eq:AveCon})-\eqref{eq:one-shot} as DPAC-OSP algorithm in short.
In the following, we present the  accuracy-privacy trade-offs of the DPAC-OSP algorithm with Laplace and Gaussian noises, both widely used in the literature to achieve  $\epsilon$-differential privacy and $(\epsilon,\delta)$-differential privacy,  respectively.

\vspace{1em}
\begin{proposition}[Trade-offs of DPAC-OSP algorithm] 
Consider the DPAC-OSP algorithm \eqref{eq:AveCon}-\eqref{eq:one-shot}. 
\begin{itemize}
\item {\bf Laplace Mechanism}. For any $\epsilon> 0$, $\mu>0$, and $\xi_i\sim\mathcal{L}(0,\sigma_\xi)$, if $\epsilon$-differential privacy of $\db$ under $\mu$-adjacency is preserved, then there must hold 
\begin{eqnarray}
    &\sigma_\xi \geq {\mu}/{\epsilon} \label{eq:sig-xi-L}\,,\\
    &\lim_{t\rightarrow\infty} \mathbb{E} |x_i(t)-d^\ast|^2 \geq {2\mu^2}/({n\epsilon^2}) \label{eq:error-G}\,.
\end{eqnarray}
\item {\bf Gaussian Mechanism}. For any $\epsilon\geq 0$, $\delta\in(0,1)$, $\mu>0$, and $\xi_i\sim\mathcal{N}(0,\sigma_\xi^2)$, if the $(\epsilon,\delta)$-differential privacy of  $\db$ under $\mu$-adjacency is preserved, then there must hold 
\begin{eqnarray}
    &\sigma_\xi\geq {\mu}/{{\kappa^{-1}_\epsilon}(\delta)} \label{eq:sig-xi-G}\,,\\
    &\lim_{t\rightarrow\infty} \mathbb{E} |x_i(t)-d^\ast|^2  \geq \mu^2/[n{\kappa^{-2}_\epsilon}(\delta)]\,\label{eq:error-L}\,.
\end{eqnarray}
\end{itemize}
\end{proposition}

By denoting $\xib=\col(\xi_1,\ldots,\xi_n)$, the mechanism for  privacy analysis is given by $\mathscr{M}(\db)=\db+\xib$, from which  \eqref{eq:sig-xi-L} and \eqref{eq:sig-xi-G} can be easily verified by recalling \cite{le2013differentially} and \cite{balle2018improving,wang2023differential}, respectively. As for \eqref{eq:error-G} and \eqref{eq:error-L}, they are clear by noting that each agent state converges to the averaged state $d^\ast+\mathbf{1}_n^\top\xib/n$ \cite{mesbahi2010graph} and then using \eqref{eq:sig-xi-L} and \eqref{eq:sig-xi-G}.

If all private data $\{d_i\}$ is stored in a center, then the average can be computed in a centralized way as $\sum_{i=1}^n d_i/n$. For differential privacy concern, the center generates a random noise $\xi$ and publishes the  \emph{perturbed} average as 
\begin{equation}\label{eq:cent}
x = \frac{1}{n}\mathbf{1}_n^\top\db + \xi.
\end{equation}
For convenience, we call the above Differentially Private Centralized Averaging algorithm  (\ref{eq:cent}) as DPCA algorithm in short. The resulting accuracy-privacy trade-offs under Laplace and Gaussian mechanisms are given below.

\vspace{1em}
\begin{proposition}[Trade-offs of DPCA algorithm] 
Consider the DPCA algorithm \eqref{eq:cent}.
\begin{itemize}
\item {\bf Laplace Mechanism}. For any ${\epsilon>0}$, $\mu>0$, and $\xi\sim\mathcal{L}(0,\sigma_\xi)$, if the $\epsilon$-differential privacy of $\db$ under $\mu$-adjacency is preserved, then there holds
$$\mathbb{E} |x-d^\ast|^2 = 2\sigma_\xi^2 \geq {2\mu^2}/{(n^2 \epsilon^2)}.
$$
\item {\bf Gaussian Mechanism}. For any $\epsilon\geq 0$, $\delta\in(0,1)$, $\mu>0$, and $\xi\sim\mathcal{N}(0,\sigma_\xi^2)$, if the $(\epsilon,\delta)$-differential privacy of $\db$ under $\mu$-adjacency is preserved, then there holds 
$$\mathbb{E} |x-d^\ast|^2 = \sigma_\xi^2 \geq {\mu^2}/[{n^2{\kappa^{-2}_\epsilon}(\delta)]}.
$$
\end{itemize}
\end{proposition}

The proof of the above proposition is clear  by using \cite[Theorem 2]{le2013differentially} and  \cite[Theorem 8]{balle2018improving} and is thus omitted.

{\bf Problem Statement.} It can be seen from Propositions 1 and 2 that under both Laplace and Gaussian mechanisms achieving the given differential privacy requirements, the achievable mean-square computation accuracy of the DPCA algorithm is  $n$ times smaller than that of the DPAC-OSP algorithm, which makes a significant difference when the network size is very large. Motivated by this gap, this paper aims to propose new differentially private average consensus algorithms which improve the accuracy-privacy trade-off in the sense of reducing and even eliminating  this gap.

\vspace{-1em}
\section{Distributed Shuffling Mechanism}
\vspace{-1em}
In this section, inspired by \cite{ruan2019secure}, we propose a distributed shuffling mechanism by employing the technique of Paillier cryptosystem \cite{paillier1999public} to generate correlated randomness in a distributed and secure manner, which will play a significant role in the initialization step of the average consensus algorithm.

Denote the local encryption and decryption operations based on the Paillier cryptosystem as $\mathrm{E}_i(\cdot)$ and $\mathrm{D}_i(\cdot)$, respectively for agent $i\in\mV$. The proposed distributed shuffling mechanism is given in Algorithm 1.

%, denoted by $\textnormal{DisShuf}(\mathbf{d};\etab\sim\calN(0,\sigma_\eta^2\Ib_n))$
\begin{algorithm}[H]
\leftline{{\bf Input:} Data $d_i$, public and private key pairs $(k_{pi},k_{si})$,}

\leftline{\qquad \quad  and a large positive integer $\bar a>>1$.}
\begin{itemize}
  \item[1.] Each agent $i\in\mV$ generates an i.i.d. noise $\eta_i$ with some probability distribution function $f_i$ and adds to the local data $d_i$: $d_i\rightarrow \bar d_i:=d_i+\eta_i$;
  \item[2.] Each agent $i\in\mV$ encrypts $-\bar d_i$ with the local public key $k_{pi}: -\bar d_i\rightarrow\mathrm{E}_i(-\bar d_i)$, and sends the local ciphertext $\mathrm{E}_i(-\bar d_i)$ and public key $k_{pi}$ to neighboring agents $j\in\mN_i$;
  \item[3.] Each agent $i\in\mV$ encrypts the noisy data $\bar d_i$ with the received public key $k_{pj}: \bar d_i\rightarrow\mathrm{E}_j(\bar d_i)$ for $j\in\mN_i$, and computes  $c_{ij} = \mathrm{E}_j(\bar d_i)\mathrm{E}_j(-\bar d_j)$ for $j\in\mN_i$;
  \item[4.] Each agent $i\in\mV$ independently and randomly generates a set of positive integers $a_{i\rightarrow j}\in[\bar a/\sqrt{2},\bar a]$, $j\in\mN_i$, and computes $(c_{ij})^{a_{i\rightarrow j}}$,  for $j\in\mN_i$;
  \item[5.] Each agent $i\in\mV$ sends the computed $(c_{ij})^{a_{i\rightarrow j}}$ to agent $j\in\mN_i$, and decrypts the received $(c_{ji})^{a_{j\rightarrow i}}$ with the local private key $k_{si}: (c_{ji})^{a_{j\rightarrow i}}\rightarrow \mathrm{D}_i((c_{ji})^{a_{j\rightarrow i}})$, $j\in\mN_i$;
  \item[6.] Each agent $i\in\mV$ multiplies each $\mathrm{D}_i((c_{ji})^{a_{j\rightarrow i}})$ by $a_{i\rightarrow j}$, $j\in\mN_i$, and computes the sum 
  $$\Delta_{i}= \sum_{j\in\mN_i}a_{i\rightarrow j}\mathrm{D}_i((c_{ji})^{a_{j\rightarrow i}})\,.$$
\end{itemize}
\leftline{{\bf Output:} $\Delta_{i}$, $i\in\mV$.}
\caption{Distributed Shuffling ($\textnormal{DiShuf}$) Mechanism}
\end{algorithm}

\vspace{1em}
\begin{remark}
The distributed shuffling process in Algorithm 1 follows the communication framework proposed in \cite{ruan2019secure} to guarantee that the communicated messages are ciphertexts, ensuring security of the actual messages and thus the related sensitive data against the eavesdropper. More explicitly, in  \cite{ruan2019secure}, {a similar computation process is implemented where node state updates iteratively with no noise added to node state}, leading to a \emph{secure} average consensus algorithm with convergence to the exact average. This is different from our cases (see the subsequent $\textnormal{DiShuf}$-based average consensus algorithms), where Algorithm 1 is implemented for only one time at the initialization step. As a result, in contrast with the average consensus algorithm in  \cite{ruan2019secure}, our $\textnormal{DiShuf}$-based average consensus algorithms need less computational and communication costs, but  at the price of sacrificing some privacy and accuracy (can see Theorems 1-6 subsequently). 
\end{remark}

It is noted that the Paillier cryptosystem has the following two significant properties: 
\begin{itemize}
    \item \emph{Homomorphic addition} 
    \begin{equation}
        \mathrm{E}_i(m_1+m_2) = \mathrm{E}_i(m_1)\mathrm{E}_i(m_2).
    \end{equation}
    \item \emph{Homomorphic multiplication} 
    \begin{equation}
        \mathrm{E}_i(km)  = \mathrm{E}_i(m)^k \,,\quad \forall k\in\mathbb{Z}^+.
    \end{equation}
\end{itemize}
Bearing in mind the above properties,  we observe that 
\[\ba{rcl}
\Delta_{i} &=& \sum_{j\in\mN_i}a_{i\rightarrow j}\mathrm{D}_i(c_{ji}^{a_{j\rightarrow i}})\,\\
&=& \sum_{j\in\mN_i}a_{i\rightarrow j}\mathrm{D}_i\big((\mathrm{E}_i(\bar d_j) \mathrm{E}_i(-\bar d_i))^{a_{j\rightarrow i}}\big) \\
&=& \sum_{j\in\mN_i}a_{i\rightarrow j}\mathrm{D}_i\big(( \mathrm{E}_i(\bar d_j-\bar d_i))^{a_{j\rightarrow i}}\big) \\
&=& \sum_{j\in\mN_i}a_{i\rightarrow j}\mathrm{D}_i( \mathrm{E}_i(a_{j\rightarrow i}(\bar d_j-\bar d_i))) \\
&=& \sum_{j\in\mN_i}a_{i\rightarrow j}a_{j\rightarrow i}(\bar d_j-\bar d_i) \\
&=& \sum_{j\in\mN_i}a_{i\rightarrow j}a_{j\rightarrow i}( d_j-d_i + \eta_j-\eta_i)\,.
\ea\]
Note that  the outputs of the $\textnormal{DiShuf}$ mechanism $\Delta_{i}$, $i\in\mV$ are correlated random variables, satisfying $\sum_{i=1}^n\Delta_{i}=0$.

In the proposed $\textnormal{DiShuf}$ mechanism, it is worth noting that by observing the communication messages, e.g., the local ciphertexts $\mathrm{E}_i(-\bar d_i)$ and $c_{ij}^{a_{i\rightarrow j}}$, the eavesdroppers have no access to the actual information, e.g., $-\bar d_i$ or $a_{j\rightarrow i}(\bar d_j-\bar d_i)$ due to the lack of the private keys. Thus, throughout the paper we assume that the ``actual" communication messages (e.g., $-\bar d_i$ and $a_{j\rightarrow i}(\bar d_j-\bar d_i)$) are secure, and will not be incorporated into the eavesdropped information in the subsequent differential privacy analysis. Moreover, if there is a malicious agent $i$,  the received $\mathrm{E}_j(-\bar d_j)$ and ${(c_{ji})^{a_{j\rightarrow i}}}$, $j\in\mathrm{N}_i$ cannot be used to infer  $\bar d_j$ as the private key $k_{sj}$ and $a_{j\rightarrow i}$ are unknown to agent $i$.

\vspace{0.5em}
\begin{remark}
Note that the Paillier cryptosystem in Algorithm 1  works on integers, while real world agent states $d_i$ and added noises $\eta_i$ are typically represented by floating point numbers in modern computing
architectures. To handle such an issue, one may multiply $\bar d_i$ by a large integer $C$ and take the nearest integer to encrypt, then divide the decrypted result by $C$ after applying the Dishuf Mechanism. By choosing a sufficiently large $C$, quantization errors can be made negligible, as addressed in \cite{ruan2019secure}.
\end{remark}

{
%\vspace{1em}
\begin{remark}
   We also stress that the main objective of the $\textnormal{DiShuf}$ mechanism is to generate correlated \emph{zero-sum} randomness $\Delta_i$. To achieve the zero-sum property, it requires to exchange messages (see Steps 2 and 5 in Algorithm 1) between each pair of nodes, implying that an undirected graph is needed. As for the extension to more general communication graph, it remains open and is of definite interest for future research.
\end{remark}
}

\vspace{-1em}
\section{$\textnormal{DiShuf}$-based Average Consensus with $(\epsilon,\delta)$-Differential Privacy: Gaussian Mechanism}
\vspace{-1em}
In this section, the $\textnormal{DiShuf}$ mechanism in Algorithm 1 is employed to develop a new $(\epsilon,\delta)$-differentially private average consensus algorithm, where the added noises follow the Gaussian distribution, leading to the Gaussian mechanism for $(\epsilon,\delta)$-differential privacy analysis.
\vspace{-1em}
\subsection{Algorithm}
\vspace{-1em}
We propose the  $\textnormal{DiShuf}$-based average consensus algorithm with Gaussian noises in Algorithm 2.

\begin{algorithm}[H]
\leftline{{\bf Input:} Data $d_i$, public and private key pairs $(k_{pi},k_{si})$,}

\leftline{\qquad \quad  $ \sigma_\eta, \sigma_\gamma>0$ and a large positive integer $\bar a>>1$.}

\begin{itemize}
  \item[1.] Each agent $i\in\mV$ implements the $\textnormal{DiShuf}$ mechanism with   $\eta_{i}\sim\calN(0,\sigma_\eta^2)$, and outputs  $\Delta_i$; 
  \item[2.] Each agent $i\in\mV$ initializes 
  \begin{equation}
      x_i(0) = d_i + \zeta\Delta_{i} +\gamma_i\,
  \end{equation}
  with ${\zeta=\frac{1}{n\bar a^2 +1}}$, and i.i.d.  Gaussian noise $\gamma_i\sim\calN(0,\sigma_\gamma^2)$;
  \item[3.] For $t=0,1,...$, run
  \begin{itemize}
      \item[3.1] Each agent $i\in\mV$ sends the local state $x_i(t)$ to the neighboring agents;
      \item[3.2] Each agent $i\in\mV$ updates its state following \eqref{eq:AveCon}.
  \end{itemize}

\end{itemize}

\caption{$\textnormal{DiShuf}$-based Average Consensus}
\vspace{-0mm}
\end{algorithm}

As $\sum_{i=1}^n\Delta_{i}=0$, it can be easily verified  that 
\[
\sum_{i=1}^nx_i(0) = \sum_{i=1}^n d_i + \sum_{i=1}^n \gamma_i\,,
\]
which is independent of $\eta_i$. In other words, the noises $\eta_i$ do not affect the average of $x_i(0)$, $i\in\mV$.
We also note that if $x_i(0)$, $i\in\mV$  are published, the corresponding differential privacy is determined by both noises  $\eta_i, \gamma_i$. In contrast with the idea of adding the i.i.d. noise $\xi_i$ to the local data $d_i$ directly as in Proposition 1, Algorithm 2 provides an extra design freedom (i.e., $\eta_i$, or $\sigma_\eta$), which affects the achievable differential privacy by publishing $x_i(0)$, $i\in\mV$, but with no influence to their average.
As a result,   a better trade-off between the privacy and the accuracy can be achieved by appropriately designing $\sigma_\eta$ and $\sigma_\gamma$.

\vspace{-1em}
\subsection{Privacy and Accuracy Analysis}
\vspace{-1em}
The differential privacy and computation accuracy of  Algorithm 2 are summarized in the following theorems, with the proof given in Appendices A and B.

\vspace{1em}
\begin{theorem}[{\bf Differential Privacy}]
For any $\epsilon\geq 0$, $\delta\in(0,1)$ and $\mu>0$, Algorithm 2 preserves  $(\epsilon,\delta)$-differential privacy of $\bf d$ under $\mu$-adjacency if 
\begin{equation}\label{eq:delta}
    \frac{1}{n\sigma_{\gamma}^2} + \frac{(n-1)\alpha^2}{\sigma_\gamma^2 + (1-\alpha)^2\sigma_\eta^2} \leq \frac{{\kappa^{-2}_\epsilon}(\delta)}{\mu^2}
\end{equation}
where 
\begin{equation}\label{eq:alpha}
    \alpha = \Big(1-\frac{1}{(2(n+\bar a^{-2}))^{n-1}}\Big)^{1/(n-1)}\,.
\end{equation}
\end{theorem}

\vspace{1em}
\begin{theorem}[{\bf Convergence}] The followings hold.
\begin{itemize}
    \item[i)] {{$|x_i(t)-x(\infty)|=\mathcal{O}(\beta^t)$}} with 
    $$x(\infty):= d^\ast +\mathbf{1}_n^\top\gammab/n\,.$$ 
    \item[ii)] $\lim_{t\rightarrow\infty} \mathbb{E} [x_i(t)-d^\ast] =0$.
    \item[iii)] $\lim_{t\rightarrow\infty} \mathbb{E} |x_i(t)-d^\ast|^2 =\sigma_{\gamma}^2/n$.
\end{itemize}
\end{theorem}

From Theorems 1 and 2, it is clear that there is a trade-off between the differential privacy level and mean-square computation accuracy. More explicitly, let 
\begin{equation}\label{eq:sigma_gamma}
    \sigma_\gamma= \frac{(1+g)\mu}{\sqrt{n} {\kappa^{-1}_\epsilon}(\delta)}\,
\end{equation}
with  $g>0$ a design freedom.
Then (\ref{eq:delta}) is satisfied if and only if there holds
\begin{equation}\label{eq:sigma_eta}\ba{rcl}
    \sigma_\eta^2 \geq \frac{(n-1)\alpha^2}{(1-\alpha)^2[{\kappa^{-1}_\epsilon}(\delta)]^2}\left[\frac{(1+g)^2\mu^2}{(1+g)^2-1} - \frac{(1+g)^2\mu^2}{n(n-1)\alpha^2}\right]\,.
\ea\end{equation}
Therefore, by Theorems 1 and 2 we can easily conclude the following result on the accuracy-privacy trade-off.

\vspace{1em}
\begin{theorem}[{\bf Trade-off}]
Let $\epsilon\geq0,\delta\in(0,\,1),\mu>0$ be any expected differential privacy levels. By choosing the noise levels $\sigma_\gamma$ and $\sigma_\eta$ satisfying (\ref{eq:sigma_gamma}) and (\ref{eq:sigma_eta}), respectively for any $g>0$,  Algorithm 2 preserves the $(\epsilon,\delta)$-differential privacy of $\bf d$ under $\mu$-adjacency, while rendering the mean-square computation error to satisfy
\begin{equation}
    \lim_{t\rightarrow\infty} \mathbb{E} |x_i(t)-d^\ast|^2 \geq {(1+g)^2 \mu^2}/{[n^2{\kappa^{-2}_\epsilon}(\delta)]}\,,
\end{equation}
where ``$=$" holds if ``$=$" in (\ref{eq:sigma_eta}) holds.
\end{theorem}

It is clear from Theorem 3 that the trade-off between the differential privacy and  mean-square computation accuracy cannot be removed, which is consistent with other differentially private average consensus algorithms \cite{nozari2017differentially,liu2020differentially}. However, we note that the achievable mean-square computation accuracy by Algorithm 2 is \emph{inversely proportional} to the \emph{square} of the agent number $n$, as the centralized Gaussian mechanism in Proposition 2. 
Moreover, by Proposition 1 when achieving the same levels of differential privacy,  the best mean-square computation error of the DPAC-OSP algorithm is $\mu^2/[n({\kappa^{-2}_\epsilon}(\delta))]$, which is $n/(1+g)^2$ larger than that of our proposed Algorithm 2 with $g>0$ an arbitrarily chosen constant. By Theorem 3 and Proposition 2, our achieved mean-square computation error is $(1+g)^2$ larger than that of the DPCA algorithm in Proposition 2. This means that the trade-off gap between the proposed Algorithm 2 and the DPCA algorithm can be almost eliminated by selecting $g>0$ small enough. {Note that a smaller $g>0$ implies a larger $\sigma_\eta$ by \eqref{eq:sigma_eta}, i.e., the variance of noise $\eta_i$ becomes larger. This in turn may bring extra computation burden  and lead the initial states $x_i(0)$ to be more dispersed.}

\vspace{-1em}
\section{$\textnormal{DiShuf}$-based Average Consensus with $\epsilon$-Differential Privacy: Laplace Mechanism}
\vspace{-1em}

In this section, the $\textnormal{DiShuf}$ mechanism in Algorithm 1 is employed to develop a new differentially private average consensus algorithm, where the noises follow the Laplace distribution, leading to the Laplace mechanism for $\epsilon$-differential privacy analysis.
\vspace{-1em}
\subsection{Algorithm}
\vspace{-1em}
As there is no guarantee in general that the sum of multiple Laplace noises follows the Laplace distribution,  Algorithm 2 with Gaussian mechanism cannot be directly adapted  to the case with Laplace mechanism for $\epsilon$-differential privacy, by replacing the Gaussian distribution by the Laplace distribution for the added noises.  In view of this, we assume that there is a secure and pre-defined agent $k^\ast\in\mV$, and propose the  $\textnormal{DiShuf}$-based average consensus algorithm with Laplace noises in Algorithm 3.

\begin{algorithm}[H]
\leftline{{\bf Input:} Data $d_i$, public and private key pairs $(k_{pi},k_{si})$}

\leftline{\qquad \quad   $\sigma_\eta, \sigma_\gamma>0$ and a large positive integer $\bar a>>1$.}

\begin{itemize}
  \item[1.] Each agent $i\in\mV$ implements the $\textnormal{DiShuf}$ mechanism with i.i.d.  Laplace noise $\eta_i\sim\calL(0,\sigma_\eta)$, and outputs $\Delta_{i}$;
  \item[2.] Each agent $i\in\mV$  initializes
  \begin{equation}
      x_i(0) = \left\{\ba{l} d_i + \zeta\Delta_{i}\,,\qquad \qquad i\in\mV/\{k^\ast\} \\ d_i+\zeta\Delta_{i} +\gamma_{i} ,\qquad i = k^\ast\ea  \right.\,
  \end{equation}
  with $\zeta{=\frac{1}{n\bar a^2 +1}}$, and i.i.d.  Laplace noise $\gamma_{k^\ast}\sim\calL(0,\sigma_\gamma)$  generated by agent $k^\ast$;
  \item[3.] For $t=0,1,...,T-1$, run
  \begin{itemize}
      \item[3.1] Each agent $i\in\mV$ sends the local state $x_i(t)$ to the neighboring agents;
      \item[3.2] Each agent $i\in\mV$ updates its state following \eqref{eq:AveCon}.
  \end{itemize}
%\leftline{{\bf Output:} $x_i(T)$, $i\in\mV$.}
\end{itemize}

%\leftline{{\bf Output:} $x_i(T)$, $i\in\mV$.}
\caption{$\textnormal{DiShuf}$-based Average Consensus}
\vspace{-0mm}
\end{algorithm}

%{\color{blue} Similar to Gaussian Mechanism, the noise $\gamma$ can not be removed as well.}

\begin{remark}
    {In both Algorithms 2 and 3, the injection of noises $\gamma_i$ is necessary. Particularly, we have $\sum_{i=1}^n x_i(0) = \sum_{i=1}^n d_i + \sum_{i=1}^n \gamma_i$ for Algorithm 2 and $\sum_{i=1}^n x_i(0) = \sum_{i=1}^n d_i + \gamma_{k^\ast}$ for Algorithm 3. This implies that the injection of noises $\gamma_i$ guarantees the differential privacy of $\db$ if the sum of initial states is published. In other words, when the initial states are published, the $\gamma_i$ provides privacy guarantees over the spanned subspace by $\mathbf{1}_n$ and thus cannot be removed. }
\end{remark}

\vspace{-1em}
\subsection{Privacy and Accuracy Analysis}
\vspace{-1em}
The differential privacy and computation accuracy of  Algorithm 3 are summarized in the following theorems.

\vspace{1em}
\begin{theorem}[{\bf Differential Privacy}]
For any $\epsilon> 0$ and $\mu>0$, Algorithm 3 preserves  $\epsilon$-differential privacy of $\mathbf{d}$ under $\mu$-adjacency if there hold
\begin{equation}\label{DP:DSLap}
\sigma_\gamma \geq h\frac{\mu}{\epsilon} \,,\quad 
\sigma_\eta\geq\frac{2\mu hn\sqrt{n-1}}{(1-\alpha)(h-1)\epsilon},
\end{equation}
 for any $h>1$, where $\alpha$ is given in \eqref{eq:alpha}.
\end{theorem}

\vspace{1em}
\begin{theorem}[{\bf Convergence}] The followings hold.
\begin{itemize}
    \item[i)] {{$|x_i(t)-x(\infty)|=\mathcal{O}(\beta^t)$}} with 
    $$x(\infty):= d^\ast +\gamma_{k^\ast}/n\,.$$ 
    \item[ii)] $\lim_{t\rightarrow\infty} \mathbb{E} [x_i(t)-d^\ast] =0$.
    \item[iii)] $\lim_{t\rightarrow\infty} \mathbb{E} |x_i(t)-d^\ast|^2 =2\sigma_{\gamma}^2/n^2$.
\end{itemize}
\end{theorem}

The proof of Theorem 4 is given in Appendix C, while for the proof of Theorem 5 it follows the same arguments  of Theorem 2 and is thus omitted for simplicity.
From Theorems 4 and 5, it is clear that there is a trade-off between the differential privacy level and mean-square computation accuracy, as in \cite{nozari2017differentially,liu2020differentially}. 

\vspace{1em}
\begin{remark}
    As shown in the proof of Theorem 4, the corresponding mechanism $\mathscr{M}(\db)$ for  privacy analysis is of $n$ dimensions and $n+1$ correlated Laplace noises. This indeed brings the corresponding privacy analysis nontrivial, since there is no guarantee that the correlation of multiple Laplace noises  still follows the Laplace distribution. To handle this issue, we first analyze a modified mechanism without considering the effect of the noise $\eta_{k^\ast}$ and then employ the resilience property of differential privacy to post-processing and the composition theorem \cite{dwork2014algorithmic} to conclude the differential privacy of $\mathscr{M}(\db)$. This, from a different perspective, means that the conditions in \eqref{DP:DSLap} are conservative and not necessary.
\end{remark}

\vspace{1em}
\begin{theorem}[{\bf Trade-off}]
Let $\epsilon>0,\mu>0$ be any expected differential privacy levels. By choosing the noise levels $\sigma_\gamma$ and $\sigma_\eta$ as in \eqref{DP:DSLap} for any $h>1$, the proposed Algorithm 3 preserves  $\epsilon$-differential privacy of ${\bf d}$ under $\mu$-adjacency, while rendering the mean-square computation error to satisfy
\begin{equation}
    \lim_{t\rightarrow\infty} \mathbb{E} |x_i(t)-d^\ast|^2 \geq \frac{2h^2\mu^2}{n^2\epsilon^2}\,,
\end{equation}
where ``$=$" holds if ``$=$" in the upper of \eqref{DP:DSLap} holds.
\end{theorem}

As concluded after Theorem 3, Theorem 6 implies that under the same level of differential privacy the resulting mean-square computation error is $h^2/n$ times smaller than that of the DPAC-OSP algorithm in Proposition 1 under Laplace mechanism, and $h^2$ larger than that of the DPCA algorithm in Proposition 2. This means that the trade-off gap between the proposed Algorithm 3 and the DPCA algorithm can be almost eliminated under Laplace mechanism by choosing $h>1$  close enough to $1$. 
{However, this will lead  the lower bound of $\sigma_\eta$ (and thus the variance of noise $\eta_i$) to be very large  by the latter of \eqref{DP:DSLap}. As a result, this may bring extra computation burden  and lead the initial states $x_i(0)$ to be more dispersed.}
%{\color{blue} Similar to $g$ in Gaussian Mechanism, with $h$ closer to 1, corresponding noise $\eta$ becomes larger, leading to computation cost.}

% \subsection{Privacy Against Malicious Agents}
% For $i\in\mV/\{k^\ast\}$, if agent $i$ connects only to a malicious agent, its local data $d_i$ would be accurately determined by the malicious agent since it does not add a noise in Algorithm 3, step 2. Then the differential privacy is destroyed. Thus avoiding single neighbor connecting is of great significance.
%Results in this section is the same with that in Section IV by using \cite[Theorem 3, 4]{ruan2019secure}, thus this section is omitted.

{
\vspace{1em}
\begin{remark}
In contrast, Algorithm 3 is able to achieve $\epsilon$-differential privacy by injecting Laplace noises, but at the price of requiring a secure and pre-defined agent $k^\ast$, while Algorithm 2 is established on Gaussian mechanism by injecting Gaussian noises for preserving a looser $(\epsilon,\delta)$-differential privacy and with no need of any pre-defined agent.

\end{remark}
}

{
\begin{remark}\label{rem-ext}
In this paper, we mainly focus on  $\textnormal{DiShuf}$-based average consensus algorithms with differential privacy guarantees, by injecting commonly used  Gaussian noises and Laplace noises. We note that there are also other kinds of noises, such as truncated Laplace noise\cite{BoundedNoiseDP} and staircase-shaped noise \cite{he2020differential}, which can also be applied  to  achieve $(\epsilon,\delta)$-differential privacy and  $\epsilon$-differential privacy, respectively. The corresponding $\textnormal{DiShuf}$-based average consensus algorithm can be easily obtained by following the framework of Algorithm 3 and directly replacing the Laplace noises by the corresponding noises. 
%However, because the distribution of the sum of truncated Laplace noises or staircase-shaped noises is generally not preserved,  like Laplace noises, one can the resulting algorithm can follows the framework of  Algorithm 3 where a secure and pre-defined agent $k^\ast$ is needed.
\end{remark}
}

\vspace{-1em}
\section{Case Studies}
\vspace{-1em}
In this section, numerical examples are presented to illustrate the effectiveness of the proposed Dishuf-based average consensus algorithms.

\begin{figure}
    \centering
    \includegraphics[width=0.5\textwidth]{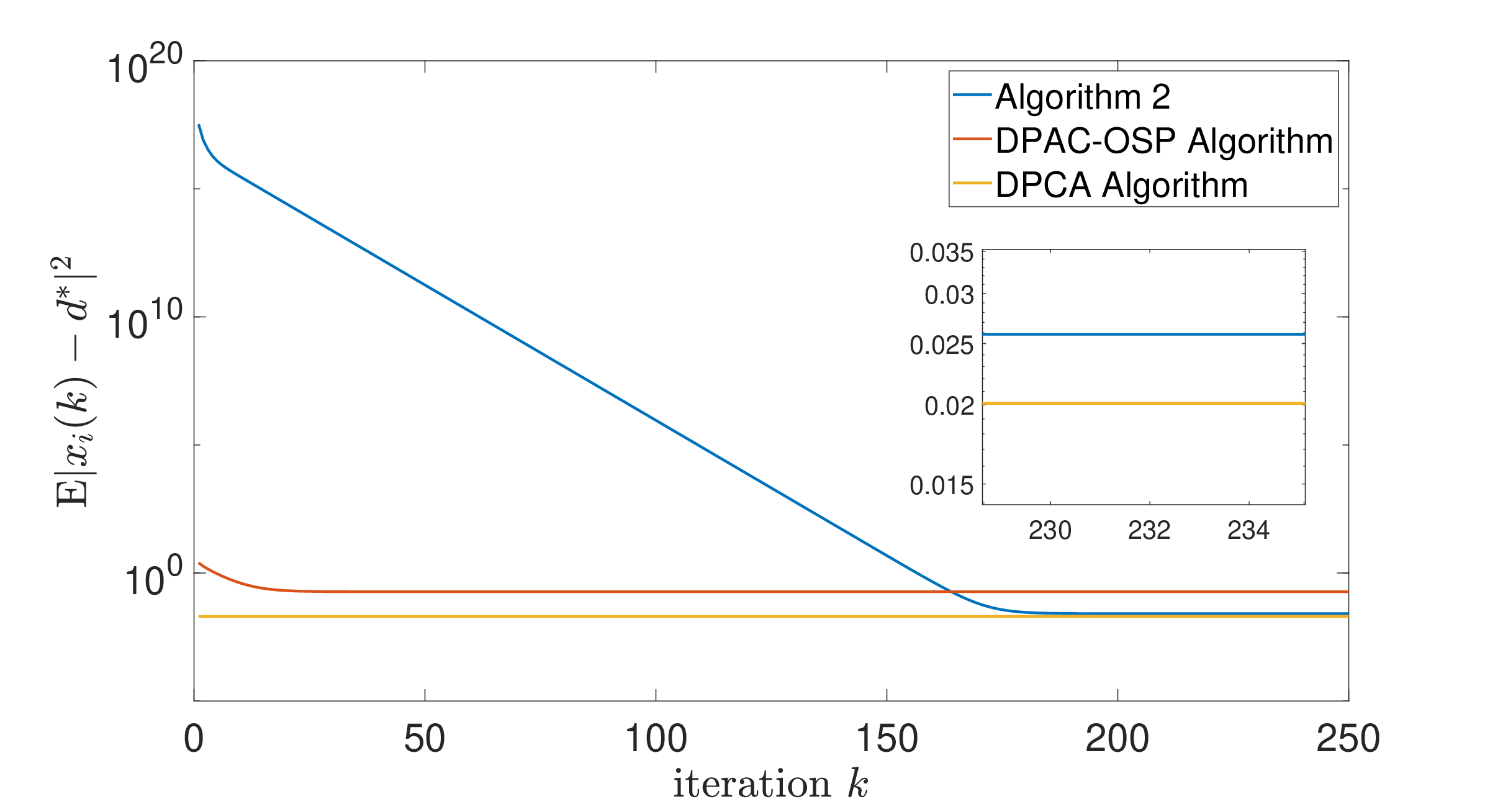}
    \caption{Trajectories of mean-square computation errors of Algorithm 2, the DPAC-OSP algorithm (\ref{eq:AveCon})-\eqref{eq:one-shot} \cite{nozari2017differentially}, and the  DPCA algorithm \eqref{eq:cent} with 200 samples}
    \label{fig:Gau}
\end{figure}

We consider  a cycle communication graph, where the agent size $n=10$ and each edge weight is assigned as $0.3$, and randomly choose the privacy-sensitive local data $d_i$  with the average  $d^\ast=13.1336$. 
Given the privacy budgets $(\epsilon, \delta, \mu)=(10,10^{-1},5)$, we let $\bar a=10^4$ and implement Algorithm 2 with Gaussian noise level $\sigma_{\gamma}=0.4500$ and $\sigma_\eta$  satisfying \eqref{eq:sigma_eta} with an appropriate $g>0$. By decreasing $g$ from 3 to 0.01, it is  shown in Table \ref{tab:Gau} that  the resulting  mean-square computation accuracy $\sum_{i=1}^n\mathbb{E} |x_i(\infty)-d^\ast|^2/n$  decreases to that of the DPCA algorithm \eqref{eq:cent}, i.e., the accuracy-privacy trade-off gap decreases as $g$ decreases. Moreover, as shown in Figure \ref{fig:Gau}, under $g=0.01$ the resulting mean-square computation accuracy is smaller than that of the DPAC-OSP algorithm (\ref{eq:AveCon})-\eqref{eq:one-shot} \cite{nozari2017differentially}, and almost recover that of the DPCA algorithm \eqref{eq:cent}.

% \begin{table}
%     \centering
%     \caption{Sensitive local data $d_i$}
%     \begin{tabular}{c|c|c|c|c|c}
%     \hline
%         $i$ & 1 & 2 & 3 & 4 & 5\\
%         \hline
%         $d_i$ & 14.3018 & 10.2806 & 11.1176 & 10.2264 & 13.7550\\
%         \hline
%         $i$ & 6 & 7 & 8 & 9 & 10\\
%         \hline
%         $d_i$ & 13.4903 & 14.5939 & 14.5446 & 14.0276 & 14.9984\\
%         \hline
%     \end{tabular}
%     \label{tab:d}
% \end{table}

\begin{table}
    \centering
    \caption{Mean-square computation accuracy  of Algorithm 2 with different $g$'s in contrast with the DPCA algorithm \eqref{eq:cent}}
    \begin{tabular}{c|c|c|c|c||c}
    \hline
        g & 3 & 2 & 1 & 0.01 & DPCA\\
        \hline
        Accuracy & 0.3234 & 0.1694 & 0.0827 & 0.0259 & 0.0201\\
        \hline
    \end{tabular}
    \label{tab:Gau}
\end{table}

\begin{figure}
    \centering
    \includegraphics[width=0.5\textwidth]{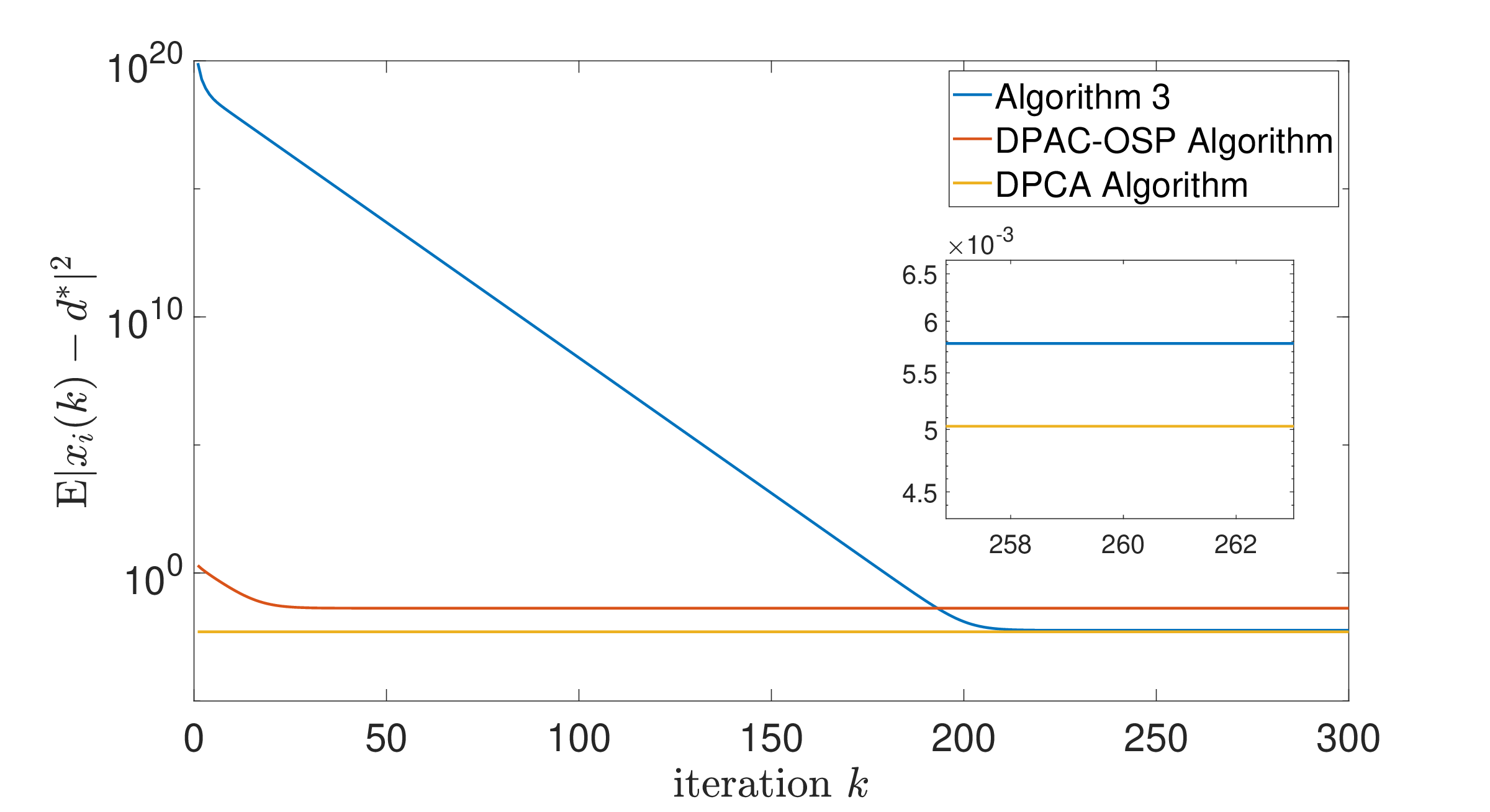}
    \caption{Trajectories of  mean-square computation errors of Algorithm 3, the DPAC-OSP algorithm (\ref{eq:AveCon})-\eqref{eq:one-shot}  \cite{nozari2017differentially}, and the DPCA algorithm \eqref{eq:cent} with 200 samples}
    \label{fig:Lap}
\end{figure}

Given $\epsilon$-differential privacy with $(\epsilon, \mu)=(10,5)$, we implement Algorithm 3 with Laplace noise levels $\sigma_{\gamma},\sigma_{\eta}$  satisfying \eqref{DP:DSLap} with an appropriate $h>1$. Similarly, 
 it is shown in Table \ref{tab:Lap} that the resulting  mean-square computation accuracy  decreases to that of the DPCA algorithm \eqref{eq:cent} as $h$ decreases to one, i.e., the accuracy-privacy trade-off gap decreases as $g$ decreases. Moreover, under  $h=1.1$ the resulting mean-square computation accuracy is smaller than that of the DPAC-OSP algorithm  (\ref{eq:AveCon})-\eqref{eq:one-shot} \cite{nozari2017differentially}, and almost recover that of the DPCA algorithm \eqref{eq:cent}, as shown in Figure \ref{fig:Lap}.

\begin{figure}
    \centering
    \begin{subfigure}{1\linewidth}
        \includegraphics[width=1\textwidth]{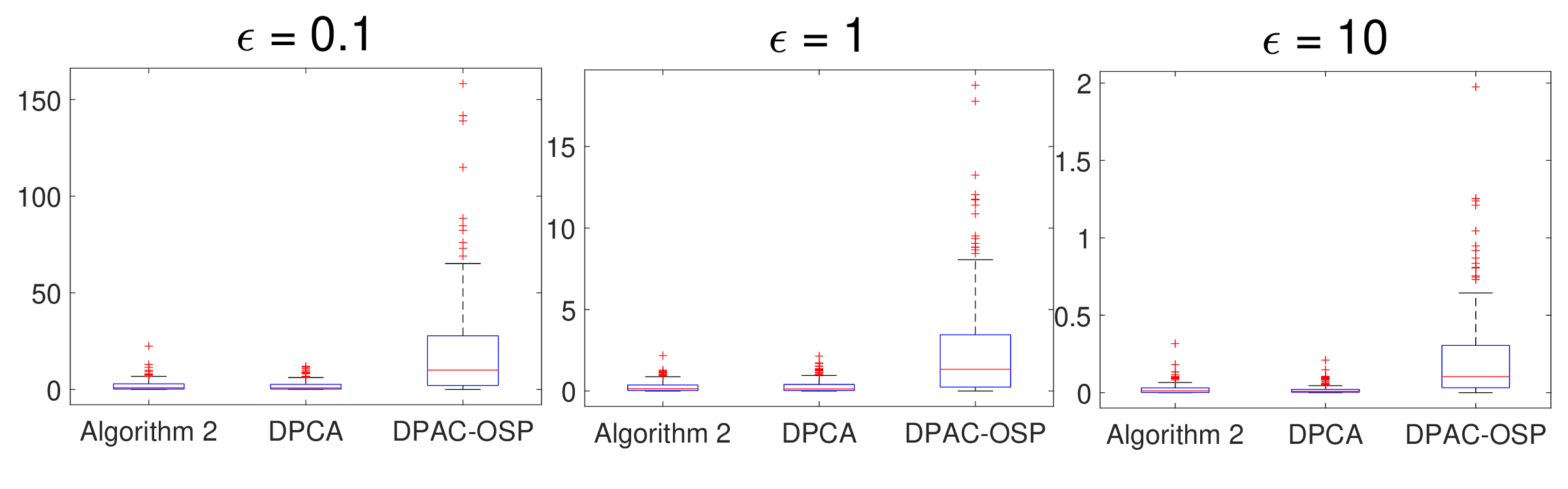}
        \caption{Gaussian Mechanism}
    \end{subfigure}
    \vfill
    \begin{subfigure}{1\linewidth}
        \includegraphics[width=1\textwidth]{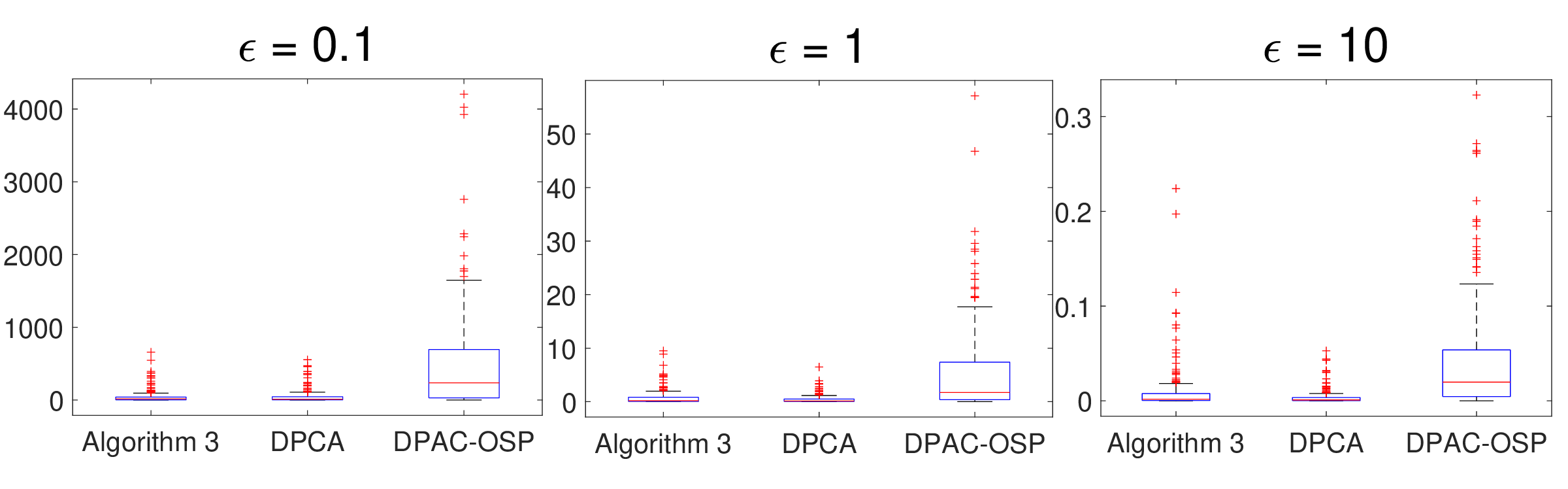}
        \caption{Laplace Mechanism}
    \end{subfigure}
    \caption{Box chart of computation errors $\frac{1}{n}\sum_{i=1}^{n}|{x_i}(\infty)-x^\ast|^2$ of our proposed algorithms, the DPCA algorithm and the DPAC-OSP algorithm under different $\epsilon$ with 200 samples}
    \label{fig:error-epsilon}
\end{figure}

\begin{figure}
    \centering
    \begin{subfigure}{1\linewidth}
        \includegraphics[width=1\textwidth]{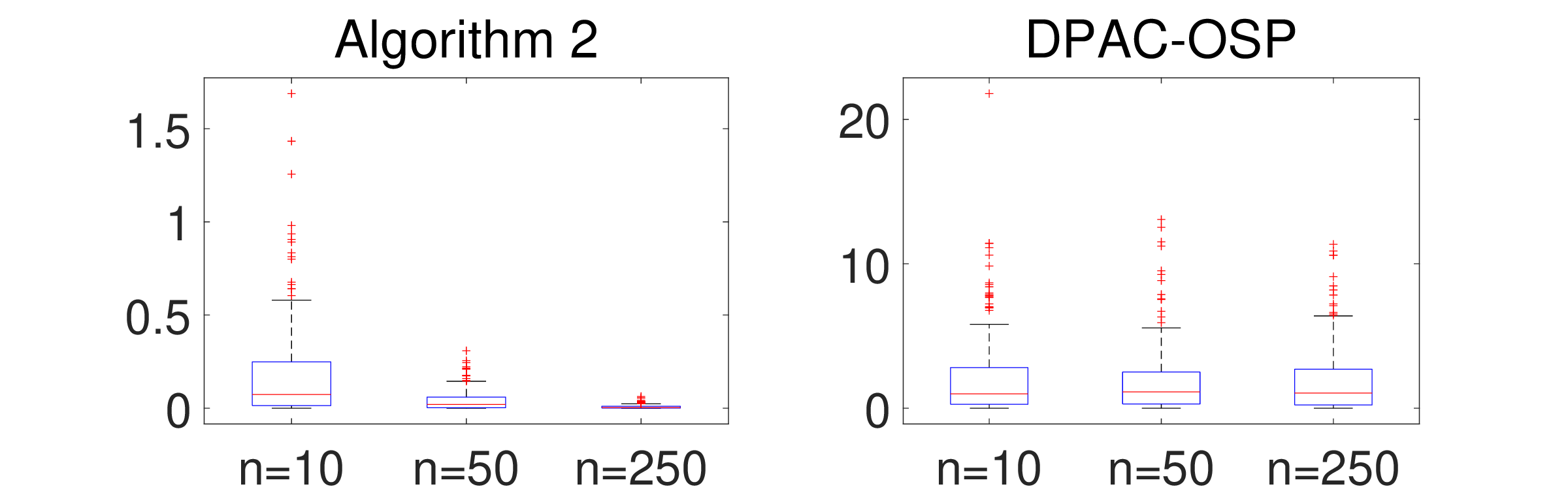}
        \caption{Gaussian Mechanism}
    \end{subfigure}
    \vfill
    \begin{subfigure}{1\linewidth}
        \includegraphics[width=1\textwidth]{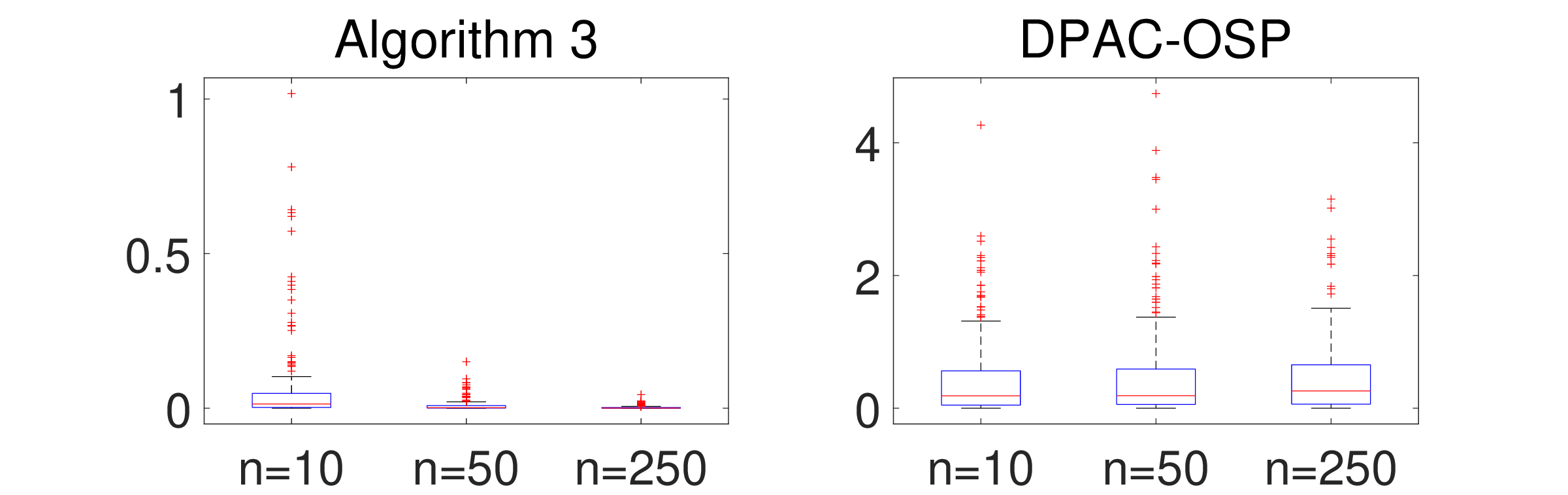}
        \caption{Laplace Mechanism}
    \end{subfigure}
    \caption{Box chart of computation errors $\sum_{i=1}^{n}|{x_i}(\infty)-x^\ast|^2$ of our proposed algorithms and the DPAC-OSP algorithm under different $n$ with 200 samples}
    \label{fig:re_e2_n}
\end{figure}

{

Moreover, we compare the computation accuracy among our algorithms, the DPCA algorithm and the DPAC-OSP algorithm under varying $\epsilon = 0.1,1,10$.  It can be seen from Fig. \ref{fig:error-epsilon} that under different privacy level $\epsilon$,  our proposed algorithms under both Gaussian and Laplace mechanisms show better computation performance and  almost recover the computation accuracy of the DPCA algoritms. 

Further,  it can be seen from Proposition 1 that the  network computation error $\sum_{i=1}^{n}|{x_i}(\infty)-x^\ast|^2$ of the DPAC-OSP algorithm is independent of the network size $n$, while  from Theorems 3 and 6  such error turns out inversely proportional to $n$ for the proposed \textnormal{DiShuf}-based algorithms, under both Gaussian and Laplace mechanisms. Thus, we also study the network computation error $\sum_{i=1}^{n}|{x_i}(\infty)-x^\ast|^2$  under varying network size $n=10,50,250$.  As seen from Fig. \ref{fig:re_e2_n}, one can observe that under both Gaussian and Laplace mechanisms, such error strictly decreases as $n$ increases for our proposed algorithms, while for the DPAC-OSP algorithm the error stays almost the same. This in turn shows that the proposed \textnormal{DiShuf}-based algorithms are more suitable for a large-scale network.

}

%\begin{table}
%    \centering
%    \begin{tabular}{c|c|c|c|c|c}
%    \hline
%        $i$ & 1 & 2 & 3 & 4 & 5\\
%        \hline
%        $d_i$ & 14.0226 & 14.3639 & 11.7126 & 10.9245 & 11.7525\\
%        \hline
%        $i$ & 6 & 7 & 8 & 9 & 10\\
%        \hline
%        $d_i$ & 11.6571 & 13.4906 & 11.5472 & 14.2565 & 14.9910\\
%        \hline
%    \end{tabular}
%    \caption{Sensitive local data $d_i$ under Laplace Mechanism}
%    \label{tab:Lap_d}
%\end{table}

\begin{table}
    \centering
    \caption{Mean-square computation accuracy of Algorithm 3 with different $h$'s in contrast with the DPCA algorithm \eqref{eq:cent}}
    \begin{tabular}{c|c|c|c|c||c}
    \hline
        h & 4 & 3 & 2 & 1.1 & DPCA\\
        \hline
        Accuracy & 0.0832 & 0.0473 & 0.0220 & 0.0058 & 0.0050\\
        \hline
    \end{tabular}
    \label{tab:Lap}
\end{table}

\vspace{-1em}
\section{Conclusion}
\vspace{-1em}
In this paper, we studied the problem of average consensus with differential privacy of initial states, for the purpose of improving the accuracy-privacy trade-off performance such that the trade-off of the centralized averaging approach with differential privacy can be (almost) recovered. To achieve such an objective, we proposed a distributed shuffling mechanism based on the Paillier cryptosystem to generate correlated zero-sum randomness. By
randomizing each local privacy-sensitive initial state with an
i.i.d. Gaussian noise and the output of the mechanism using
Gaussian noises, the resulting average consensus
algorithm was shown to be able to eliminate the gap in the sense that the accuracy-privacy trade-off of the centralized averaging approach can be (almost) recovered by adjusting a design parameter to be small enough.
We also showed that such a design framework could be extended to the one using Laplace noises with the improved privacy-accuracy trade-off preserved. {Future research works of interest include the extension to distributed optimization and cooperative control for better trade-offs under directed and time-varying communication graphs.}

%\vspace{-1em}
\appendix
\vspace{-1em}
\section{Proof of Theorem 1}
Let $a_{ij}=\zeta a_{i\rightarrow j}a_{j\rightarrow i}$ and denote $\Ab\in\mathbb{R}^{n\times n}$ by a matrix satisfying $[\Ab]_{ij}=-a_{ij}$ for $(i,j) \in \mathrm{E}$, $[\Ab]_{ij}=0$ for $(i,j) \notin \mathrm{E}$ and $[\Ab]_{ii}=\sum_{j\in\mN_i}a_{ij}$ for all $i\in\mathrm{V}$. It is clear that $\Ab$ is symmetric and positive semi-definite, and  for any $i\in\mV$, there hold
\[\ba{l}
{[\Ab]}_{ii} \leq \frac{(n-1)\bar a^2}{n\bar a^2 +1}\,,\quad |[\Ab]_{ij}| \leq \frac{\bar a^2}{n\bar a^2 +1}\,,\quad \forall j\neq i,\\
\sum_{j=1}^n {[\Ab]}_{ij} =0\,.
\ea\] 
In view of the above analysis, the matrix $\Ab$ indeed can be regarded as a Laplacian matrix of graph ${\rm G}$.
Let us arrange the eigenvalues of $\Ab$ in the increasing order as $\lambda_1^{\Ab} \leq \lambda_2^{\Ab}\leq ...\leq \lambda_n^{\Ab}$, associated with the normalized and mutually orthogonal
eigenvectors $\ub_1,\ub_2,\ldots,\ub_n$, respectively. It is clear that $\ub_1=\mathbf{1}_n/\sqrt{n}$, and $0=\lambda_1^{\Ab} < \lambda_2^{\Ab}\leq ...\leq \lambda_n^{\Ab}<2$ {by \cite[Theorem 2.8]{mesbahi2010graph} and Gershgorin Circle Theorem \cite{horn_johnson_1985}}. For convenience, we denote $\Lambda_{\Ab}=\diag\{\lambda_1^{\Ab},\lambda_2^{\Ab},\ldots,\lambda_n^{\Ab}\}$, and $\Ub=[\Ub_1;\Ub_2]$ with $\Ub_1=\ub_1$, $\Ub_2=[\ub_2;\cdots;\ub_n]$.

Next we proceed to provide a tighter lower-bound to the second eigenvalue $\lambda_2^{\Ab}$.
It is clear that the matrix $\Pb:=\Ib_n-\Ab$ is symmetric and doubly stochastic, with each element $[\Pb]_{ij}\geq \frac{\bar a^2}{2(n\bar a^2 +1)}=\frac{1}{2(n+\bar a^{-2})}$ for $j\in\mN_i\cup\{i\}$ and $[\Pb]_{ij}=0$ for $j\notin\mN_i\cup\{i\}$. According to \cite[Proposition 1]{nedic2010constrained}, one can obtain 
\[\ba{rcl}
&&|[\Pb^k]_{ij} - \frac{1}{n}| \\
\leq && 2\frac{1+{(2(n+\bar a^{-2}))^{(n-1)}}}{1-{(2(n+\bar a^{-2}))^{-(n-1)}}} \Big(1-\frac{1}{(2(n+\bar a^{-2}))^{n-1}}\Big)^{\frac{k}{n-1}}
\ea\]
for all $i,j\in\mV$ and $k\geq1$, which yields
\begin{equation}\label{eq:P}\ba{rcl}
&&\|\Pb^k - \frac{1}{n}\mathbf{1}_n\mathbf{1}_n^\top\| \\
&\leq&  2n\frac{1+{(2(n+\bar a^{-2}))^{(n-1)}}}{1-{(2(n+\bar a^{-2}))^{-(n-1)}}} \Big(1-\frac{1}{(2(n+\bar a^{-2}))^{n-1}}\Big)^{\frac{k}{n-1}}\\
&=& \mathcal{O}(\alpha^k)
\ea\end{equation}
with $\alpha<1$ defined in (\ref{eq:alpha}).
With this in mind, we note that
\[\ba{rcl}
&&\Pb^k - \frac{1}{n}\mathbf{1}_n\mathbf{1}_n^\top \\
&=&  \Ub(\Ib_n-\Lambda_{\Ab})^k \Ub^\top -\Ub_1\Ub_1^\top\\
&=& \Ub\diag\{(1-\lambda_1^\Ab)^k-1,(1-\lambda_2^\Ab)^k,\ldots,(1-\lambda_n^\Ab)^k\} \Ub^\top\\
&=& \Ub\diag\{0,(1-\lambda_2^\Ab)^k,\ldots,(1-\lambda_n^\Ab)^k\} \Ub^\top\,,\\
&=&\sum_{j=2}^n(1-\lambda_j^\Ab)^k\ub_j\ub_j^\top\,.
\ea\]
Thus, by (\ref{eq:P}) and the inequality $0 < \lambda_2^{\Ab}\leq ...\leq \lambda_n^{\Ab}<2$, we have
\[
\max\{|1-\lambda_2^\Ab|,|1-\lambda_n^\Ab|\}\leq \alpha\,.
\]
Now we consider the following two cases:
\begin{itemize}
    \item[(i)] If $\lambda_2^\Ab<1$, we have $1-\lambda_2^\Ab\leq \alpha$, implying $1>\lambda_2^\Ab\geq 1-\alpha$;
    \item[(ii)] If $\lambda_2^\Ab\geq1$, {obviously we have $\lambda_2^\Ab > 1-\alpha$}.
    % we have $\lambda_2^\Ab-1 \leq \lambda_n^\Ab-1\leq \alpha$, implying $1+\alpha \geq \lambda_2^\Ab\geq 1-\alpha$;
\end{itemize}
In view of the above two cases, we have $\lambda_2^\Ab \geq 1-\alpha$.

Bearing in mind the previous analysis, we now analyze the differential privacy of the proposed averaged consensus algorithm, and focus on the mechanism
\begin{equation}\label{AveAlg}
    \mathscr{M}(\db) = \Pb \db - \Ab \etab + \gammab.
\end{equation}

\vspace{-1em}
For Theorem 1, $\etab:=[\eta_1;\ldots;\eta_n]\sim\mathcal{N}(0,\sigma_\eta^2)^n$ and $\gammab:=[\gamma_1;\ldots;\gamma_n]\sim\mathcal{N}(0,\sigma_\gamma^2)^n$.
For any $\mathcal{M}\subseteq\mathbb{R}^n$,  then we note that
\[\ba{rcl}
&&\mathbb{P}\big(\mathscr{M}(\db)\in \mathcal{M}\big) \\
&=& \mathbb{P}\big(\Ub^\top(\Pb \db - \Ab \etab + \gammab)\in \Ub^\top\mathcal{M}\big)\\
&=&\mathbb{P}\big(((\Ib_n-\Lambda_{\Ab})\Ub^\top \db - \Lambda_{\Ab}\Ub^\top \etab + \Ub^\top\gammab)\in \Ub^\top\mathcal{M}\big) 
\ea\]
where by simple calculations we have
\[
- \Lambda_{\Ab}\Ub^\top \etab + \Ub^\top\gammab \sim \mathcal{N}(0,\Sigma^2)\,
\]
with $\Sigma = \diag(\sigma_\gamma,\sqrt{\sigma_\gamma^2+(\lambda_2^{\Ab}\sigma_\eta)^2},\ldots,\sqrt{\sigma_\gamma^2+(\lambda_n^{\Ab}\sigma_\eta)^2})$.
Thus, we define the mechanism
\[
\mathscr{M}_\dag(\db) := \Sigma^{-1}(\Ib_n-\Lambda_{\Ab})\Ub^\top \db + \omegab\,,\quad \omegab\sim\mathcal{N}(0,1)^n\,,
\]
and can see that for any $\epsilon\geq 0$, $\delta\in(0,1)$ and $\mu>0$, the inequality
\begin{equation}\label{eq:DP2}
    \mathbb{P}\big(\mathscr{M}(\db) \in \mathcal{M} \big) \leq e^\epsilon \mathbb{P}\big(\mathscr{M}(\db^\prime) \in \mathcal{M} \big) +\delta\,,\quad \forall (\db,\db')\in\textnormal{Adj}(\mu)
\end{equation}
holds for all $\mathcal{M}\subseteq\mathbb{R}^n$, if and only if for all $\mathcal{M}_\dag\subseteq\mathbb{R}^n$,
\begin{equation}\label{eq:DP3}
    \mathbb{P}\big(\mathscr{M}_\dag(\xb) \in \mathcal{M}_\dag \big) \leq e^\epsilon \mathbb{P}\big(\mathscr{M}_\dag(\xb^\prime) \in \mathcal{M}_\dag \big) +\delta
\end{equation}
holds for all $(\db,\db')\in\textnormal{Adj}(\mu)$.

Then by recalling \cite[Theorem 8]{balle2018improving}, we know that the mechanism $\mathscr{M}_\dag$ is $(\epsilon,\delta)$-differentially private if and only if 
\begin{equation}\label{eq:DP}
    \kappa_\epsilon(S_0):=\Phi(\frac{S_0}{2} - \frac{\epsilon}{S_0}) - e^{\epsilon}\Phi(-\frac{S_0}{2} - \frac{\epsilon}{S_0}) \leq \delta
\end{equation}
with the sensitivity $S_0=\max_{i\in\mV} \|\mu\Sigma^{-1}(\Ib_n-\Lambda_{\Ab})\Ub^\top\eb_i\|$.
It is further noted that
\[\ba{rcl}
S_0 &=& \max_{i\in\mV}  \sqrt{\sum_{j=1}^n\left|\frac{\mu(1-\lambda_j^{\Ab})}{\sqrt{\sigma_{\gamma}^2+\lambda_j^{\Ab}\sigma_\eta^2} }\ub_j^\top\eb_i\right|^2}\\
&\leq& \sqrt{\frac{\mu^2}{\sigma_{\gamma}^2 }\max\limits_{i\in\mV}|\ub_1^\top\eb_i|^2 + \frac{\mu^2\alpha^2}{\sigma_{\gamma}^2+(\lambda_2^{\Ab}\sigma_\eta)^2} \max\limits_{i\in\mV}\sum_{j=2}^n|\ub_j^\top\eb_i|^2}\\
&\leq& \sqrt{\frac{\mu^2}{n\sigma_{\gamma}^2 } + \frac{(n-1)\mu^2\alpha^2}{\sigma_{\gamma}^2+(1-\alpha)^2\sigma_\eta^2}}
\ea\]
where to obtain the first inequality we have used the facts that $\lambda_1^{\Ab}=0$, and the inequalities
\[\ba{l}
\max\{|1-\lambda_2^\Ab|,\ldots,|1-\lambda_n^\Ab|\}\leq \alpha\\
0 < \lambda_2^{\Ab}\leq ...\leq \lambda_n^{\Ab}\,,
\ea\]
and used the facts that $\ub_1=\mathbf{1}_n/\sqrt{n}$, $\|\ub_j\|=1$ for $j=2,\ldots,n$ and the inequality $\lambda_2^\Ab \geq 1-\alpha$  to derive the second inequality.

Thus by recalling (\ref{eq:DP}),  the mechanism $\mathscr{M}$ preserves $(\epsilon,\delta)$-differential privacy of local data $\db$ if (\ref{eq:delta}) holds.

To complete the proof, it is worth specifying the information that may be eavesdropped. At the stage of distributed shuffling,  the communication messages are encrypted and thus cannot be utilized for privacy inference by eavesdroppers due to the absence of the private keys. At the stage of average consensus, the eavesdroppers may have access to the communication messages, i.e., $\xb(t)$, $t\geq0$. It is noted that  $\xb(t)$, $t\geq 1$ can be expressed as deterministic functions of the initial states $\xb(0):=\mathscr{M}(\db)$. According to the robustness property of the differential privacy to post-processing \cite{dwork2014algorithmic}, and recalling that the mechanism $\mathscr{M}$ is $(\epsilon,\delta)$-differentially private with (\ref{eq:delta}), we can conclude that the proposed average consensus algorithm preserves $(\epsilon,\delta)$-differential privacy of $\db$ with (\ref{eq:delta}), completing the proof.

\section{Proof of Theorem 2}

{
For matrix $\Ib_n-\Lb$, denote $\{\lambda_i^{\Ib-\Lb}\}$ and $\{\boldsymbol{\nu}_i\}$ eigenvalues and corresponding normalized and mutually orthogonal eigenvectors, respectively. 
Noting that $\lambda_i^{\Lb} \in [0,2)$ and there exists only one being $0$, we have $-1 < \lambda_1^{\Ib-\Lb} \leq \dots \leq \lambda_{n-1}^{\Ib-\Lb} < \lambda_n^{\Ib-\Lb}=1$ and it can be verified that $\boldsymbol{\nu}_n = \mathbf{1}_n/\sqrt{n}$.
Then by \eqref{eq:AveCon} we have
\begin{equation}
\begin{aligned}
    \xb(t) = & (\Ib_n-\Lb)^{t}\xb(0) \\
    = & [\boldsymbol{\nu}_1; \dots; \boldsymbol{\nu}_n] \Lambda_{\Ib_n-\Lb}^t [\boldsymbol{\nu}_1; \dots; \boldsymbol{\nu}_n]^\top \xb(0) \\
    = & (\lambda_n^{\Ib-\Lb}\boldsymbol{\nu}_n\boldsymbol{\nu}_n^\top + \sum_{i=1}^{n-1} \lambda_i^{\Ib-\Lb}\boldsymbol{\nu}_i\boldsymbol{\nu}_i^\top) \xb(0) \\
    = & \frac{\mathbf{1}_n\mathbf{1}_n^\top}{n}(\db+\gammab) + \sum_{i=1}^{n-1} \lambda_i^{\Ib-\Lb}\boldsymbol{\nu}_i\boldsymbol{\nu}_i^\top \xb(0),
\end{aligned}
\end{equation}
where $\Lambda_{\Ib_n-\Lb} = \diag\{\lambda_1^{\Ib-\Lb}, \lambda_2^{\Ib-\Lb}, \ldots, \lambda_n^{\Ib-\Lb}\}$ and we used
\[
    \mathbf{1}_n^\top\xb(0)=\mathbf{1}_n^\top \db +\mathbf{1}_n^\top\gammab.
\]
Since $\max\limits_{i\in\{1,\dots,n-1\}}|\lambda_i^{\Ib-\Lb}| = \beta <1$, we have
\begin{equation}
    \lim\limits_{t\rightarrow\infty} \xb(t) = \frac{\mathbf{1}_n\mathbf{1}_n^\top}{n}(\db+\gammab),
\end{equation}
implying the node states $x_i(t)$ exponentially converge  to 
\begin{equation}\label{x_i:convg}
x_i(\infty) =  \mathbf{1}_n^\top \db/n +\mathbf{1}_n^\top\gammab/n :=x(\infty)\,
\end{equation}
with convergence rate $ln(1/\beta)$, i.e., $|x_i(t)-x(\infty)|=\mathcal{O}(\beta^t)$. This proves the statement (i).

Regarding the statements (ii) and (iii), they can be directly derived from (\ref{x_i:convg}) by taking the expectation operations.

}

\section{Proof of Theorem 4}
As in Appendix A, to analyze the differential privacy of Algorithm 3 against the eavesdropper accessing the communication messages, we consider the following mechanism
\begin{equation}
    \mathscr{M}(\db) = \Pb \db - \Ab \etab + \eb_{k^\ast}\gamma_{k^\ast}\,,
\end{equation}
where $\etab:=[\eta_1;\ldots;\eta_n]\sim\mathcal{L}(0,\sigma_\eta)^n$ and $\gamma_{k^\ast}\sim\mathcal{L}(0,\sigma_\gamma)$.

To show $\epsilon$-differential privacy of $\mathscr{M}$, we first study $\mathscr{M}_1:=\mathbf{1}_n^\top \mathscr{M}(\db)$, and observe
\[\ba{rcl}
\mathbf{1}_n^\top\mathscr{M}(\db)
&=&\mathbf{1}_n^\top \Pb\db - \mathbf{1}_n^\top \Ab\etab+\mathbf{1}_n^\top\eb_{k^\ast}\gamma_{k^\ast}\\
&=&\mathbf{1}_n^\top \Pb\db + \gamma_{k^\ast}.
\ea\]
Since $\sup_{{(\db,\db')} \in Adj(\mu)}{\|\mathbf{1}_n^\top \Pb \db-\mathbf{1}_n^\top \Pb \db'\|}_1=\mu$, by  \cite[Theorem 2]{le2013differentially}, it is clear that $\mathscr{M}_1$ is $\epsilon/h$-differentially private with 
$\sigma_\gamma$ satisfying \eqref{DP:DSLap}.

Then we observe that  $rank(\Ub_2)=n-1$ and $\mathbf{1}_n^\top\Ub_2=0$, which implies that the $k^\ast$-th row of $\Ub_2$ can be expressed by a linear combination of the remaining rows of $\Ub_2$. As a consequence, denote by $\bar \Ub_2\in\mathbb{R}^{(n-1)\times (n-1)}$ the matrix by removing the $k^\ast$-th row of $\Ub_2\in\mathbb{R}^{n\times (n-1)}$, and have $rank(\bar \Ub_2)=n-1$, i.e., $\bar \Ub_2$ is invertible.
Thus we consider 
\[\ba{rcl}
\mathscr{M}_2(\db) 
={(\bar\Ub_2^\top)}^{-1}{\mathbf{T}}^{-1}\Ub_2^\top(\Ib_n-\eb_{k^\ast}\mathbf{1}_n^\top)\Pb\db-\bar\eta, \\
\ea\]
where $\mathbf{T}=\diag\{\lambda_2^\Ab,\lambda_3^\Ab,\ldots,\lambda_n^\Ab\}\in\mathbb{R}^{(n-1)\times (n-1)}$, satisfying $\lambda_i^\Ab \geq {1-\alpha}$ for $i=2,\ldots,n-1$, and ${\bar\eta}=[\eta_1;\ldots;\eta_{k^\ast-1};\eta_{k^\ast+1};\ldots;\eta_n]\in\mathbb{R}^{n-1}$.
Note that 
\[\ba{rcl}
&& \Ub_2^\top(\Ib_n-\eb_{k^\ast}\mathbf{1}_n^\top)\Pb\\ 
&=& \bar\Ub_2^\top\bar\Pb- \Ub_2^\top\eb_{k^\ast}\mathbf{1}_{n-1}^\top\bar\Pb \\
&=& \bar\Ub_2^\top\bar\Pb + (\Ub_2^\top\mathbf{1}_{n}- \Ub_2^\top\eb_{k^\ast})\mathbf{1}_{n-1}^\top\bar\Pb \\
&=& \bar\Ub_2^\top(\Ib_{n-1}+\mathbf{1}_{n-1}\mathbf{1}_{n-1}^\top)\bar\Pb\,,
\ea\]
where the first equation is obtained by defining $\bar\Pb\in\mathbb{R}^{(n-1)\times n}$ as the matrix by removing the $k^\ast$-th row of $\Pb$, and the second is obtained by using the fact that $\Ub_2^\top\mathbf{1}_{n}=0$. This yields that
for $(\db,\db') \in Adj(\mu)$,
\[\ba{rcl}
&&{\|{(\bar\Ub_2^\top)}^{-1}{\mathbf{T}}^{-1}\Ub_2^\top(\Ib_n-\eb_{k^\ast}\mathbf{1}_n^\top)\Pb(\db-\db')\|}_1 \\
%&=&\mu\|{(\bar\Ub_2^\top)}^{-1}{\mathbf{T}}^{-1}\Ub_2^\top(\Ib_n-\eb_{k^\ast}\mathbf{1}_n^\top)\Pb\|_1 \\
&=&\|{(\bar\Ub_2^\top)}^{-1}{\mathbf{T}}^{-1}\bar\Ub_2^\top(\Ib_{n-1}+\mathbf{1}_{n-1}\mathbf{1}_{n-1}^\top)\bar\Pb(\db-\db')\|_1 \\
&\leq&\mu\|{(\bar\Ub_2^\top)}^{-1}{\mathbf{T}}^{-1}\bar\Ub_2^\top\|_1\|(\Ib_{n-1}+\mathbf{1}_{n-1}\mathbf{1}_{n-1}^\top)\|_1\|\bar\Pb\|_1 \\
&\leq&\mu\frac{\sqrt{n-1}}{1-\alpha}\|(\Ib_{n-1}+\mathbf{1}_{n-1}\mathbf{1}_{n-1}^\top)\|_1\|\bar\Pb\|_1 \\
&\leq&2\mu\frac{n\sqrt{n-1}}{1-\alpha}, \\
\ea\]
where the second inequality is obtained by  $\|{(\bar\Ub_2^\top)}^{-1}{\mathbf{T}}^{-1}\bar\Ub_2^\top\|_1\leq \sqrt{n-1}\|{(\bar\Ub_2^\top)}^{-1}{\mathbf{T}}^{-1}\bar\Ub_2^\top\|\leq\frac{\sqrt{n-1}}{1-\alpha}$, and the last inequality is obtained by using $\|(\Ib_{n-1}+\mathbf{1}_{n-1}\mathbf{1}_{n-1}^\top)\|_1 = n $ and $\|\bar\Pb\|_1\leq \|\Pb\|_1 \leq 1-\frac{\bar a^2}{2(n\bar a^2+1)} + (n-1)\frac{\bar a^2}{n\bar a^2+1} \leq 2$.

Thus, by \cite[Theorem 2]{le2013differentially}, $\mathscr{M}_2(\db)$ preserves $(1-1/h)\epsilon$-differential privacy with $\sigma_\eta$ satisfying \eqref{DP:DSLap}.
Further, $\mathscr{M}_2(\db)-\eb_r\eta_r$ preserves $(1-1/h)\epsilon$-differential privacy, as the differential privacy is resilient to post-processing \cite{dwork2014algorithmic}.

Towards this end, we note that
$\mathscr{M}_2(d) - \eb_r\eta_r={(\bar\Ub_2^\top)}^{-1} \mathbf{T}^{-1} \Ub_2^\top(\Ib_n-\eb_{k^\ast}\mathbf{1}_n^\top)\mathscr{M}(d)$. By defining
\[
\Qb = \begin{bmatrix}\mathbf{1}_n^\top \\ {(\bar\Ub_2^\top)}^{-1} \mathbf{T}^{-1} \Ub_2^\top(\Ib_n-\eb_{k^\ast}\mathbf{1}_n^\top) \end{bmatrix}\,,
\]
we then can conclude that the mechanism 
\[
\mathscr{M}_\dag(\db) := \Qb\mathscr{M}(\db)
\]
is $\epsilon$-differentially private by noting that  $\mathscr{M}_\dag(\db)$ is a composition of mechanisms $\mathscr{M}_1$ and $\mathscr{M}_2(\db)-\eb_r\eta_r$ \cite[Theorem 3.14]{dwork2014algorithmic}.
This thus completes the proof by verifying that $\Qb$ is invertible and recalling that the differential privacy is resilient to post-processing \cite{dwork2014algorithmic}.

%\subsection{Proof of Theorem 5}
%Since
%$$\mathbb{E} [\frac{1}%{n^2}\|\mathbf{1}_n\mathbf{1}_n^\top\gammab\|^2]=\sigma_{\gamma}^2,
%$$
%we have
%\[
%\lim_{t\rightarrow\infty} \mathbb{E}\|\xb(t)-\frac{1}%{n}\mathbf{1}_n\mathbf{1}_n^\top \db\|^2 = \sigma_{\gamma}^2.
%\]

%The proof is thus completed.

%\subsection{Proof of Theorem 3}
%Without loss if generality, we seek to avoid direct leakage of agent 1's %local data with some accuracy i.e., $\bar d_1 = d_1 + \eta_1$ or $d_1 + %\gamma_1$. Assume $j = 2, 3, ...\in\mN_1$.

%From the perspective of agent 1, the leakage may happen in DiShuf stage where $(c_{1j})^{a_{1\rightarrow j}}$ is sent, or in Average Consensus stage where its initial state $x_1(0) = d_1 + \zeta\Delta_{i} + \gamma_1$ is published. For the former one, $d_1$ is protected by $a_{1\rightarrow j}$, but its neighboring agent $j$ could calculate $a_{1\rightarrow j}(\bar d_1 - \bar d_j)$. Then write the latter one as $x_1(0) = (d_1 + \gamma_1) + \zeta(C_2 + C_3 + ...)$. Here $C_j = a_{j\rightarrow 1}a_{1\rightarrow j}(\bar d_j - \bar d_1)$, and is known to and only to corresponding agent $j$. Thus if agent 1 connects to more than one agents, any of its neighboring agent, including possible malicious agent, cannot calculate $(C_2 + C_3 + ...)$, and then $d_1 + \gamma_1$.

\bibliographystyle{IEEEtran}
\bibliography{Auto_DPAC}
\end{document}